\documentclass[10pt,draftclsnofoot,onecolumn,journal]{IEEEtran}
\usepackage[utf8]{inputenc}
\usepackage[pdftex]{graphicx}
\usepackage{textcomp}
\usepackage{pdfpages}
\usepackage{multirow}
\usepackage{listings}
\usepackage{float}
\usepackage[caption=false]{subfig}
\usepackage{graphics}
\usepackage{array}
\usepackage{color}
\usepackage{midfloat}
\usepackage[cmex10]{amsmath}
\usepackage{moreverb}
\usepackage{cite}
\usepackage{fixltx2e}
\usepackage{url}
\usepackage{fancyhdr}
\usepackage{rotating}
\usepackage{algorithmic}
\usepackage{algorithm}
\usepackage{paralist}
\usepackage{amssymb}
\usepackage{array}

\begin{document}

\title{Epidemic and Cascading Survivability of Complex Networks}

\author{\IEEEauthorblockN{Marc Manzano, Eusebi Calle, Jordi Ripoll, Anna Manolova Fagertun,\\ Victor Torres-Padrosa, Sakshi Pahwa, Caterina Scoglio}%
\thanks{Marc Manzano, Eusebi Calle, Jordi Ripoll and Victor Torres-Padrosa are with University of Girona, Spain. Anna Manolova Fagertun is with Technical University of Denmark, Denmark. Sakshi Pahwa and Caterina Scoglio is with Kansas State University, USA. Corresponding author: Marc Manzano (email: mmanzano@eia.udg.edu - marcmanzano@ksu.edu).}}%

\maketitle \thispagestyle{plain}

\begin{abstract}
Our society nowadays is governed by complex networks, examples being the power grids, telecommunication networks, biological networks, and social networks. It has become of paramount importance to understand and characterize the dynamic events (e.g. failures) that might happen in these complex networks. For this reason, in this paper, we propose two measures to evaluate the vulnerability of complex networks in two different dynamic multiple failure scenarios: epidemic-like and cascading failures. Firstly, we present \emph{epidemic survivability} ($ES$), a new network measure that describes the vulnerability of each node of a network under a specific epidemic intensity. Secondly, we propose \emph{cascading survivability} ($CS$), which characterizes how potentially injurious a node is according to a cascading failure scenario. Then, we show that by using the distribution of values obtained from $ES$ and $CS$ it is possible to describe the vulnerability of a given network. We consider a set of 17 different complex networks to illustrate the suitability of our proposals. Lastly, results reveal that distinct types of complex networks might react differently under the same multiple failure scenario.
\end{abstract}

\begin{IEEEkeywords}
Network Characterization, Epidemics, Cascading Failures, Multiple Failures, Complex Networks
\end{IEEEkeywords}

\section{Introduction\label{sec:intro}}
Telecommunication networks, power grids, water distribution networks, transport networks or fuel distribution networks are critical infrastructures that play a vital role in our modern society. Such crucial networks do not display regular organizations, ergo they have also been addressed as \emph{complex networks}. The study of complex networks not only comprises critical infrastructures, but also any other kind of network with non-trivial features. Social networks, biological networks, online social networks and mobile social networks \cite{Yang2012msn} are solid examples of complex networks.

Our society of nowadays is governed by complex networks. For instance, people have become more and more dependent on communication networks, either for business or leisure purposes. In addition, this dependency is expected to grow, considering the myriad of new emerging technologies and services such as smart-cities, cloud computing, e-Health, the Internet of the Things, MANETs, etc. Consequently, the period of time for which a user can operate terminals without network connectivity is becoming very short; and if a large-scale failure occurred, it would impact a significant percentage of the world's population. Another example is the online social networks such as Twitter or Facebook. In August 2013 a single tweet of a billionaire investor made Apple shares rise over \$500 \cite{tweet}, showing how a single message can spread and reach millions of users within hours. These two examples depict how important it is to understand the events that might occur on complex networks. From now on, in this work we are going to use the term \emph{failure} to refer to any event that causes disruption in the normal functioning of a complex network.

Many different protection and restoration techniques for single failures have been extensively analyzed in recent decades (e.g. see \cite{Sterbenz20101245}). Furthermore, multiple failures such as natural disasters or physical attacks have also been studied \cite{azim2012vulnerability}. According to the taxonomy introduced in \cite{manzano2013endurance}, there are two types of multiple failures. While \emph{static} multiple failures are essentially one-off failures that affect one or more elements (nodes or links) simultaneously at any given point, \emph{dynamic} failures have a temporal dimension. In this paper we consider \emph{dynamic} multiple failures, which we implement through \emph{epidemic} and \emph{cascading} failures. On one hand, an epidemic-like failure propagation occurs when, at a given time, a node or a group of them start spreading an infection. In this case the failure (e.g. infection) propagates through physical neighbors. On the other hand, cascading failures occur when a node (or a group of them) fails, and as a consequence, other parts in the network fail as well due to an overloading of the capacity. Cascading failures do not necessarily propagate through physical contact, i.e. one node failure can cause a failure to a non-adjacent node due to the network load balancing.

In contrast with single failures, in the case of multiple failures it is nonviable to define proper reactive strategies. Thus, since the reasonable approach to address such large-scale failures involves the designing phase of a network, it has become of paramount importance to define new metrics able to evaluate the vulnerability of networks in the case of multiple failure scenarios. Appropriate metrics can help network engineers and operators to detect the most critical parts of a network. Although a new generic metric suitable to accurately evaluate the robustness in static multiple failure scenarios has been recently presented in \cite{manzano2013endurance}, to the best of our knowledge there are no metrics able to evaluate the robustness under dynamic multiple failure scenarios. 

In our previous work \cite{manzano2013drcn} we presented a metric called epidemic survivability. In this paper we go one step further and we extend the work by considering broader type of failures: dynamic multiple failures. In addition, we extend the number of networks considered for testing of the failure scenarios to 17, as compared to 6 in the previous work. Consequently, here we consider 2 telecommunication networks, 2 Internet Autonomous Systems (AS) networks, 5 synthetic generated networks, 1 biological network, 3 social networks and 4 power grid networks. Our aim is to take into account a wide range of different types of complex networks, and evaluate them under dynamic multiple failure scenarios. Within this context, the main contributions of this paper are:
\begin{enumerate}
	\item a new network measure called \emph{epidemic survivability} ($ES$). This feature describes the vulnerability of each node of a network under a specific epidemic scenario. 
	\item a new network measure called \emph{cascading survivability} ($CS$), which characterizes how potentially injurious a node is according to a specific cascading failure scenario. 
\end{enumerate}

We believe that our proposals can be used by the network research community to evaluate the criticality of nodes of a network under failure propagation scenarios. In addition, our metrics can be used to amplify general recovery metrics such as \cite{Cholda2008qor}.

The remainder of this work is organized as follows: Section~\ref{sec:net} presents the set of network topologies considered in this paper. In Section~\ref{sec:epidemics} we (a) introduce the state of the art related with epidemic failures; (b) review the most well-known epidemic models; (c) present our new network measure called \emph{epidemic survivability}; and (d) show a practical example of how could our proposal be used. Then, Section~\ref{sec:cascading} (a) provides a background with respect to cascading failures; (b) presents several remarkable cascading failure models; (c) defines our new metric called \emph{cascading survivability}; and (d) illustrates how to use the metric. Finally, Section~\ref{sec:conclusions} concludes this work reviewing its main contributions and findings.
\section{Network Topologies\label{sec:net}}

\begin{figure}
 \centering
 \subfloat[\emph{abilene93}]{
   \includegraphics[width=0.22\textwidth]{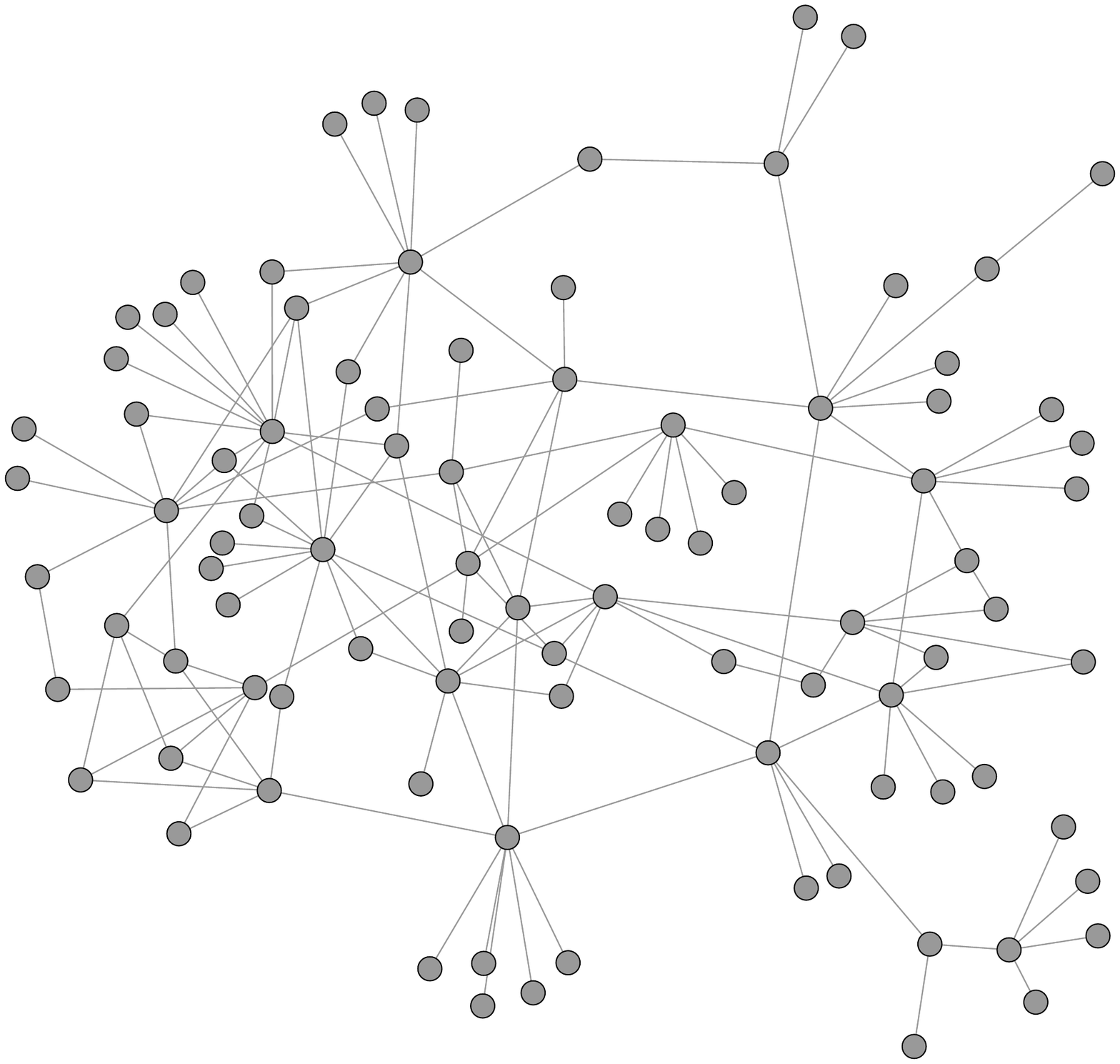}
   \label{fig:abilene93}
 }
  \subfloat[\emph{cogentco197}]{
   \includegraphics[width=0.22\textwidth]{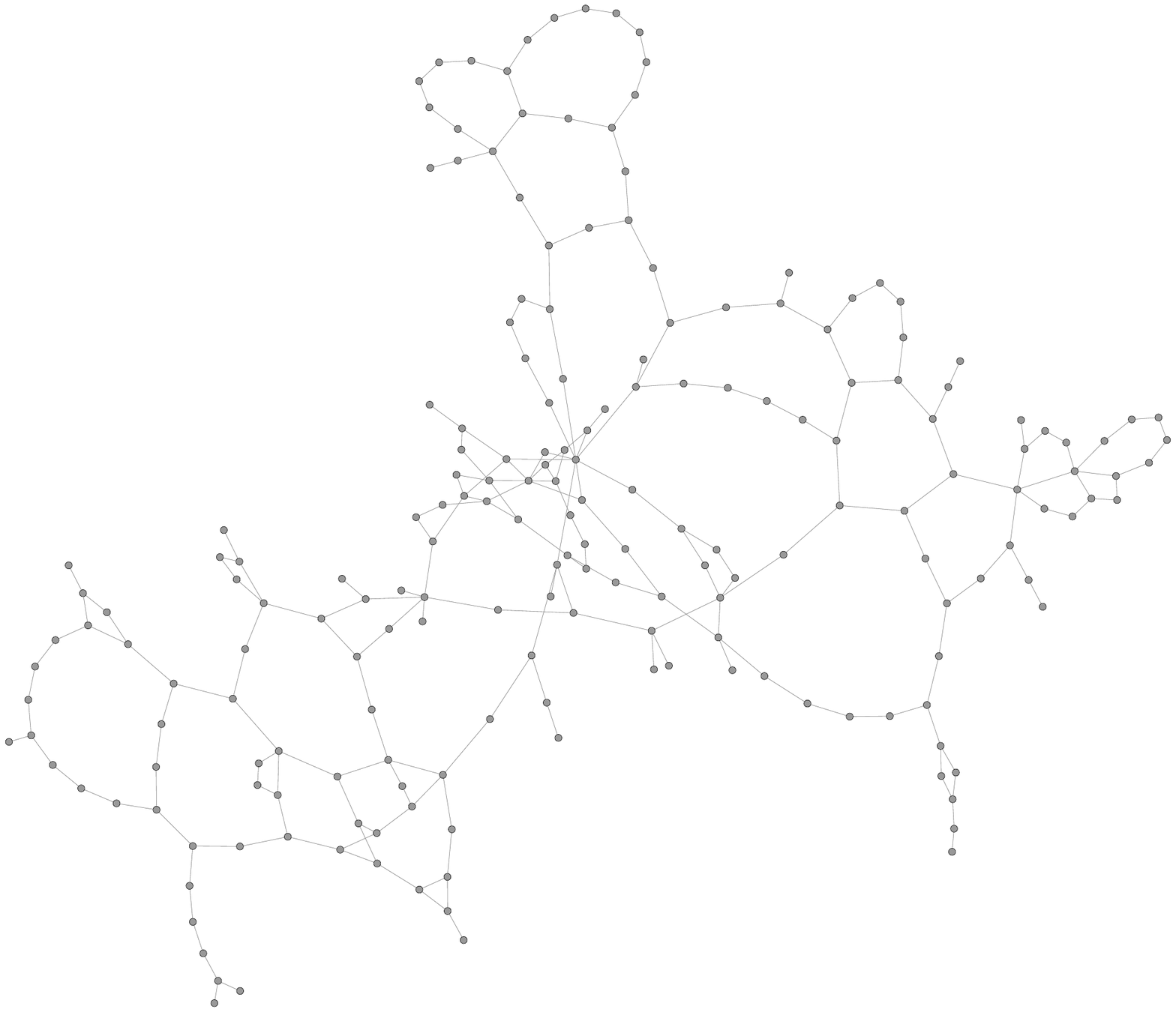}
    \label{fig:cogentco197}
  }
  \subfloat[\emph{er400}]{
    \includegraphics[width=0.22\textwidth]{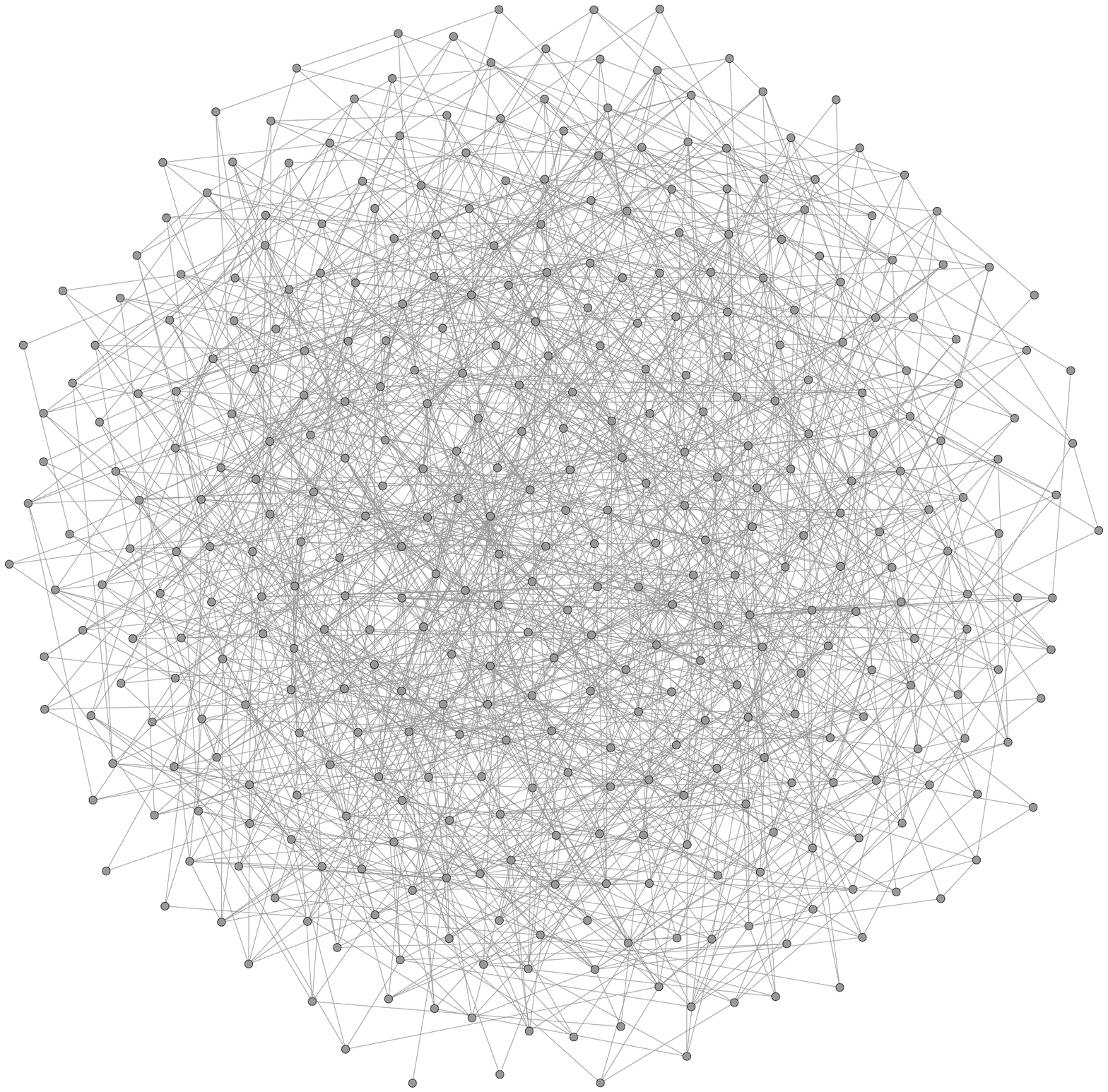}
    \label{fig:er400}
  }
  \subfloat[\emph{powerlaw400}]{
    \includegraphics[width=0.22\textwidth]{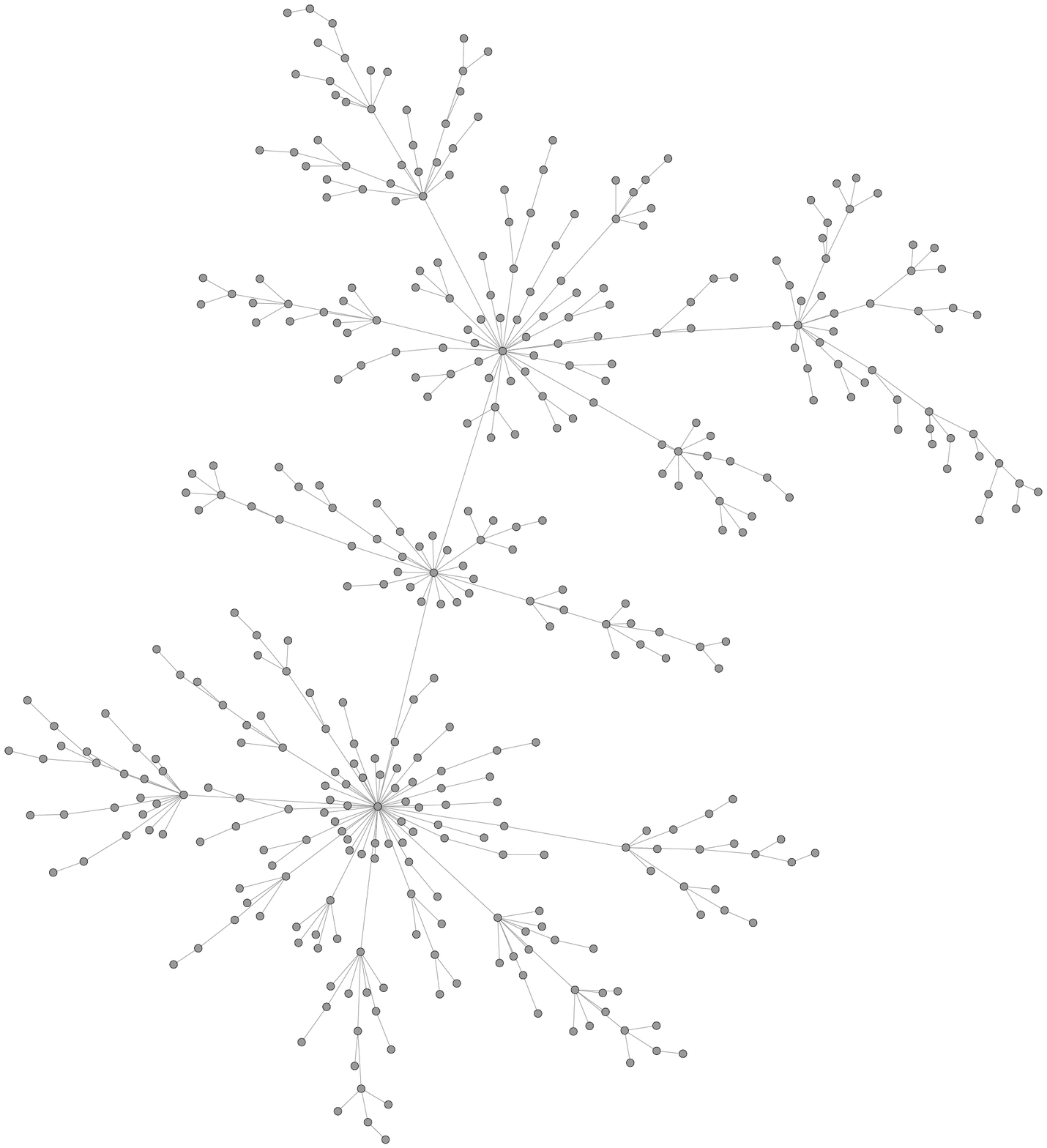}
    \label{fig:powerlaw400}
  }
\\
  \subfloat[\emph{bt400}]{
    \includegraphics[width=0.22\textwidth]{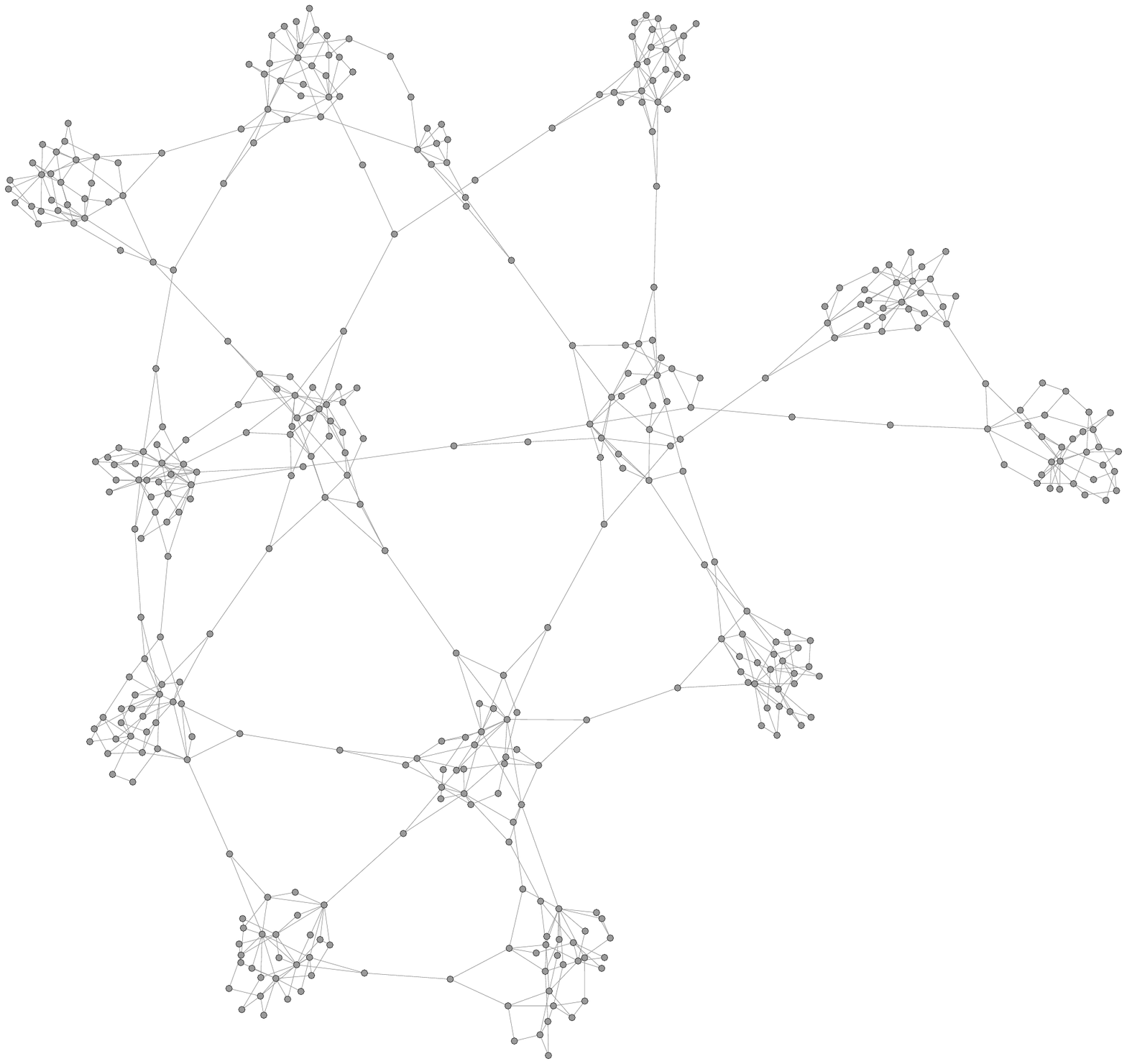}
    \label{fig:bt400}
  }
  \subfloat[\emph{homoge400}]{
    \includegraphics[width=0.22\textwidth]{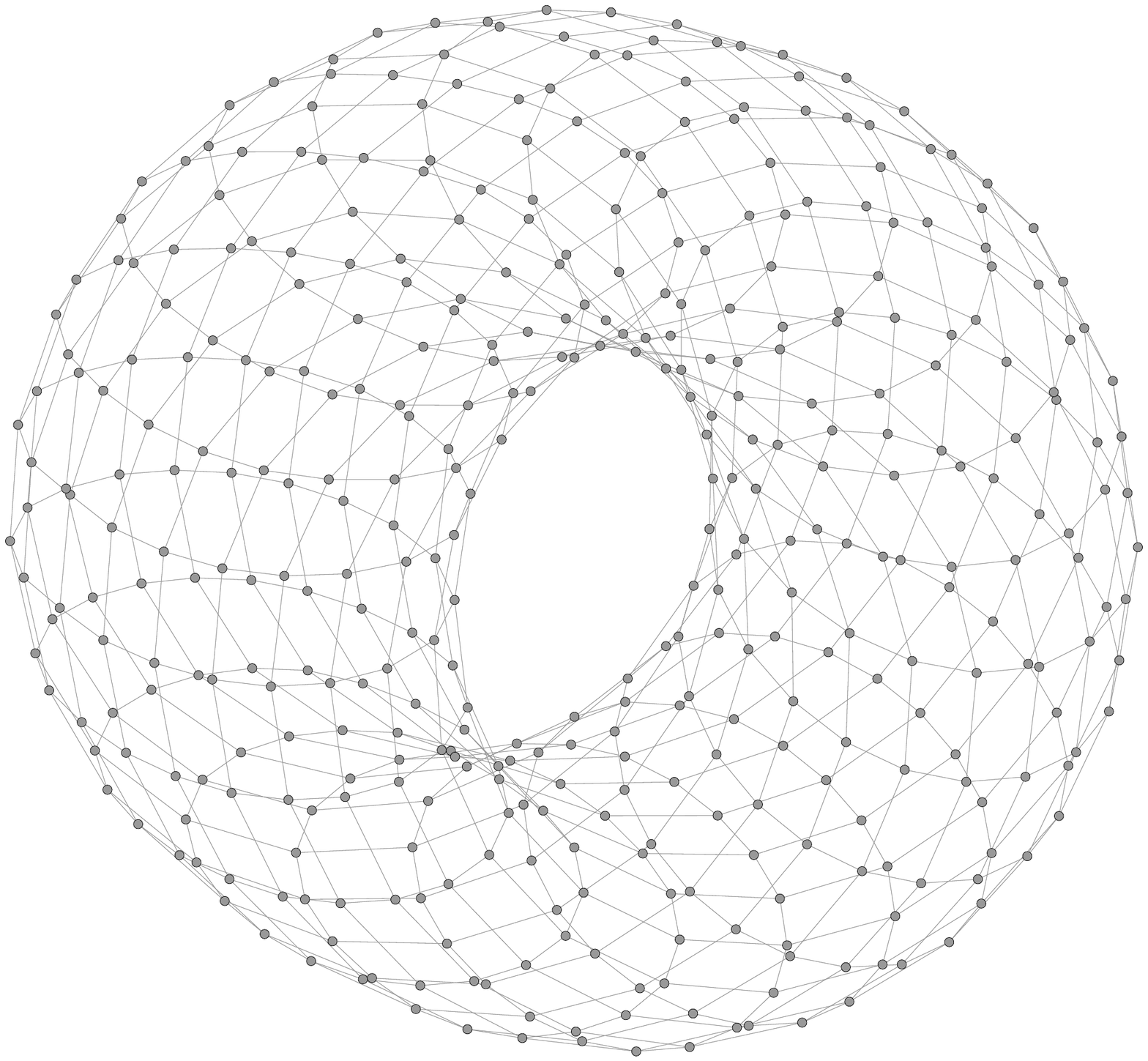}
    \label{fig:homoge400}
  }
\subfloat[\emph{bo1458}]{
  \includegraphics[width=0.22\textwidth]{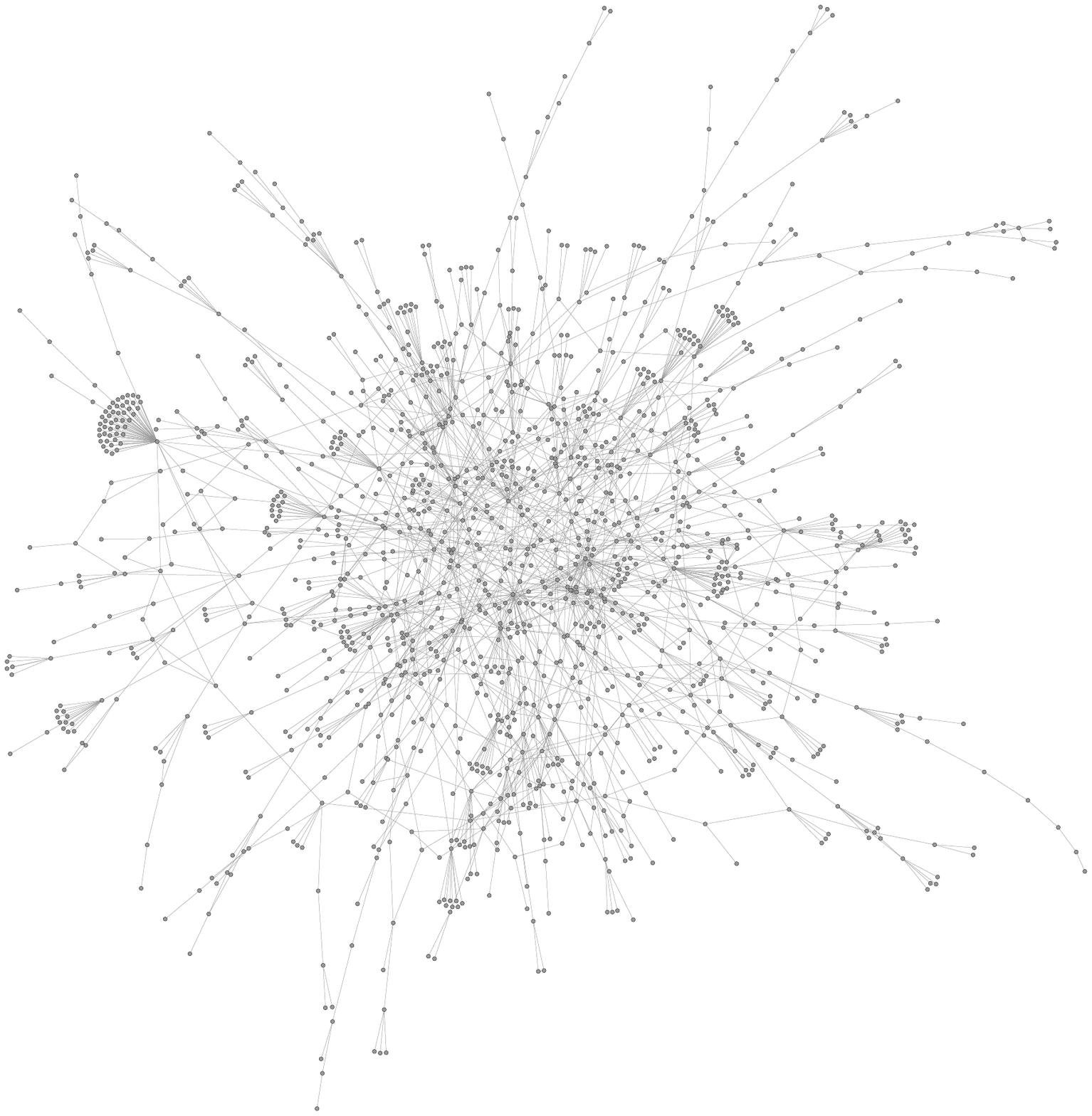}
  \label{fig:bo1458}
}
\subfloat[\emph{col4158}]{
  \includegraphics[width=0.22\textwidth]{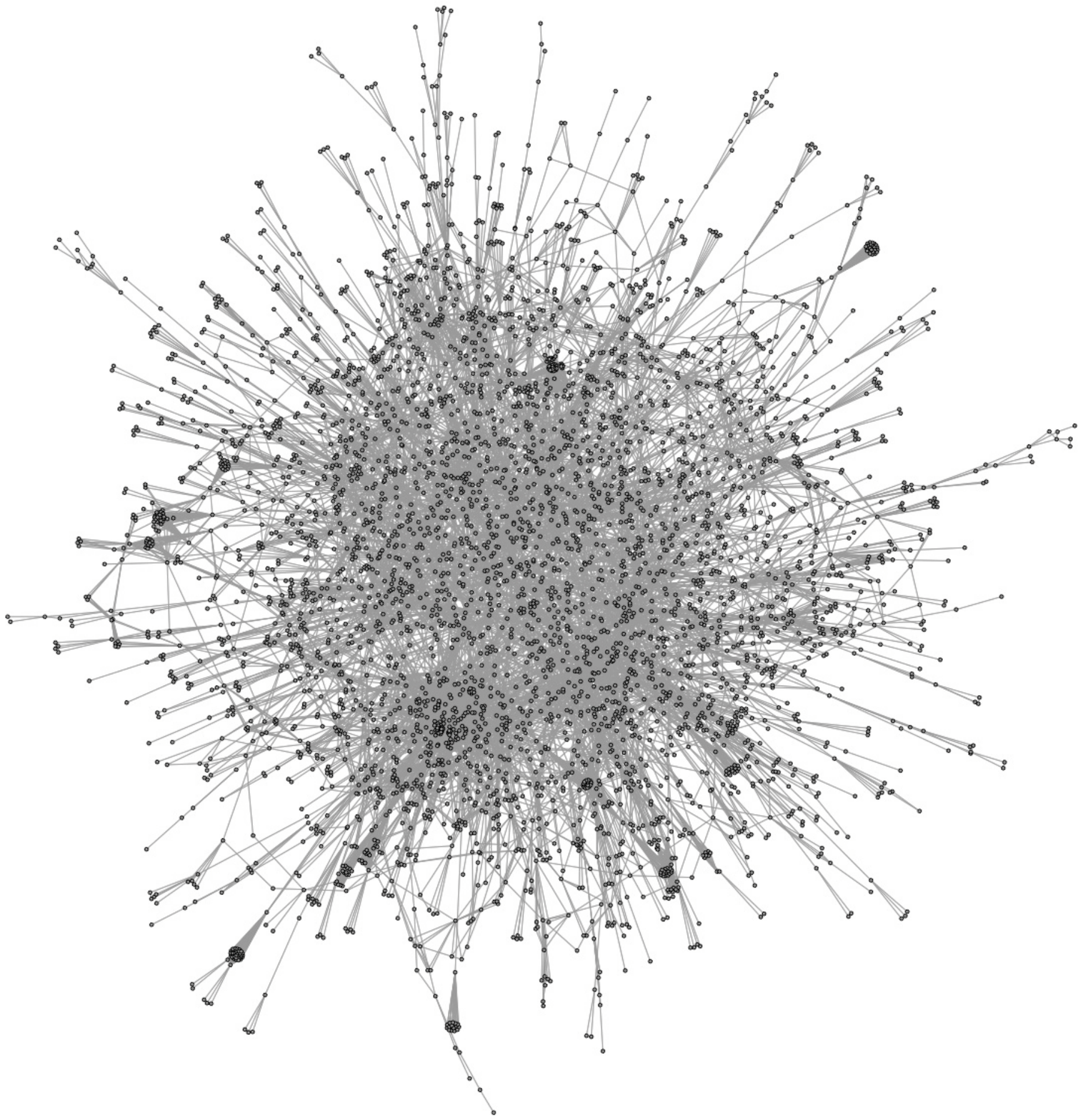}
  \label{fig:col4158}
}
  \caption{Layout of 8 out of the 17 topologies considered in this work.}
  \label{ref:top_layout}
\end{figure}

\begin{figure}
	\centering
\subfloat[\emph{col8638}]{
  \includegraphics[width=0.22\textwidth]{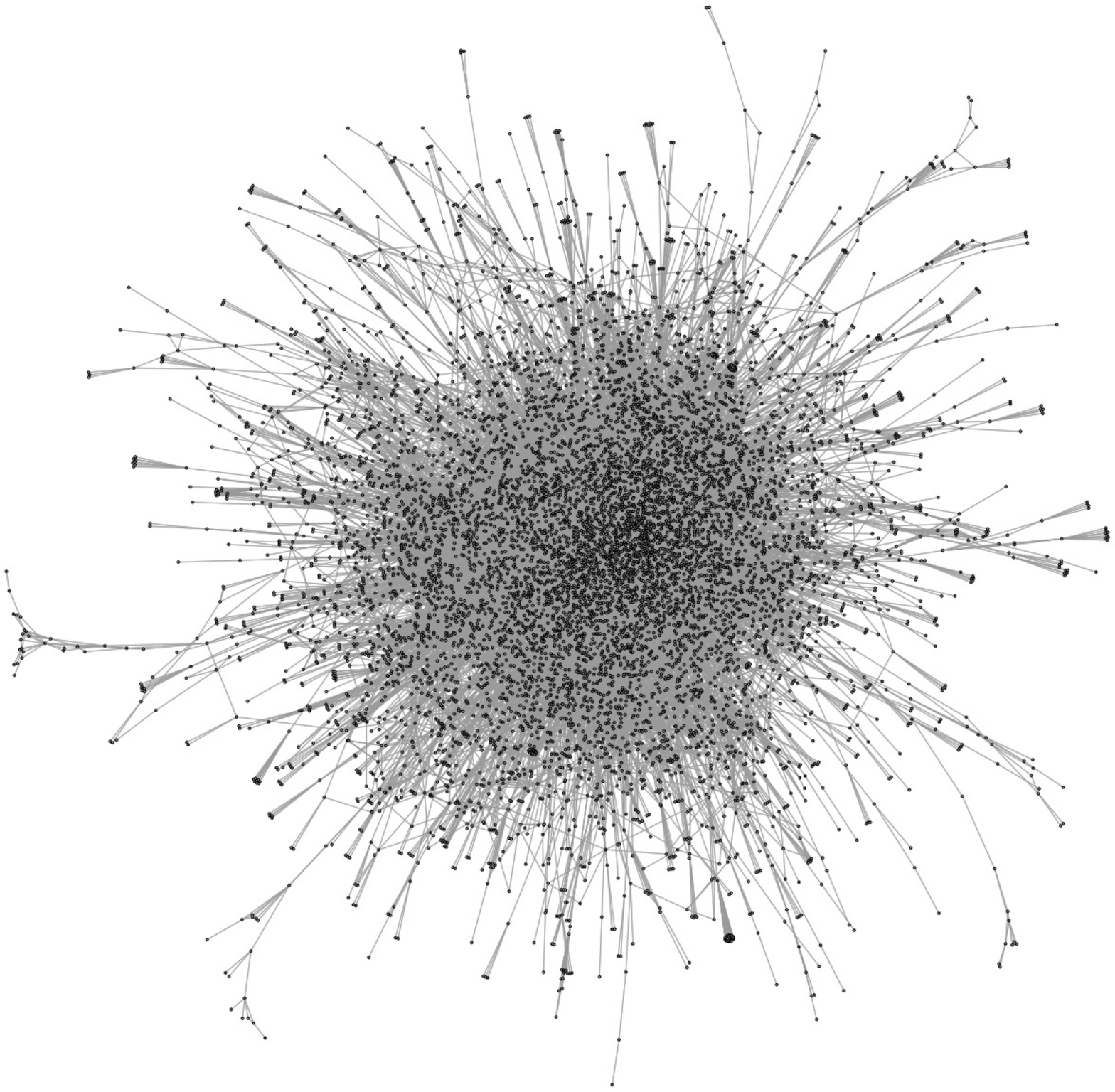}
  \label{fig:col8638}
}
\subfloat[\emph{cost37}]{
  \includegraphics[width=0.22\textwidth]{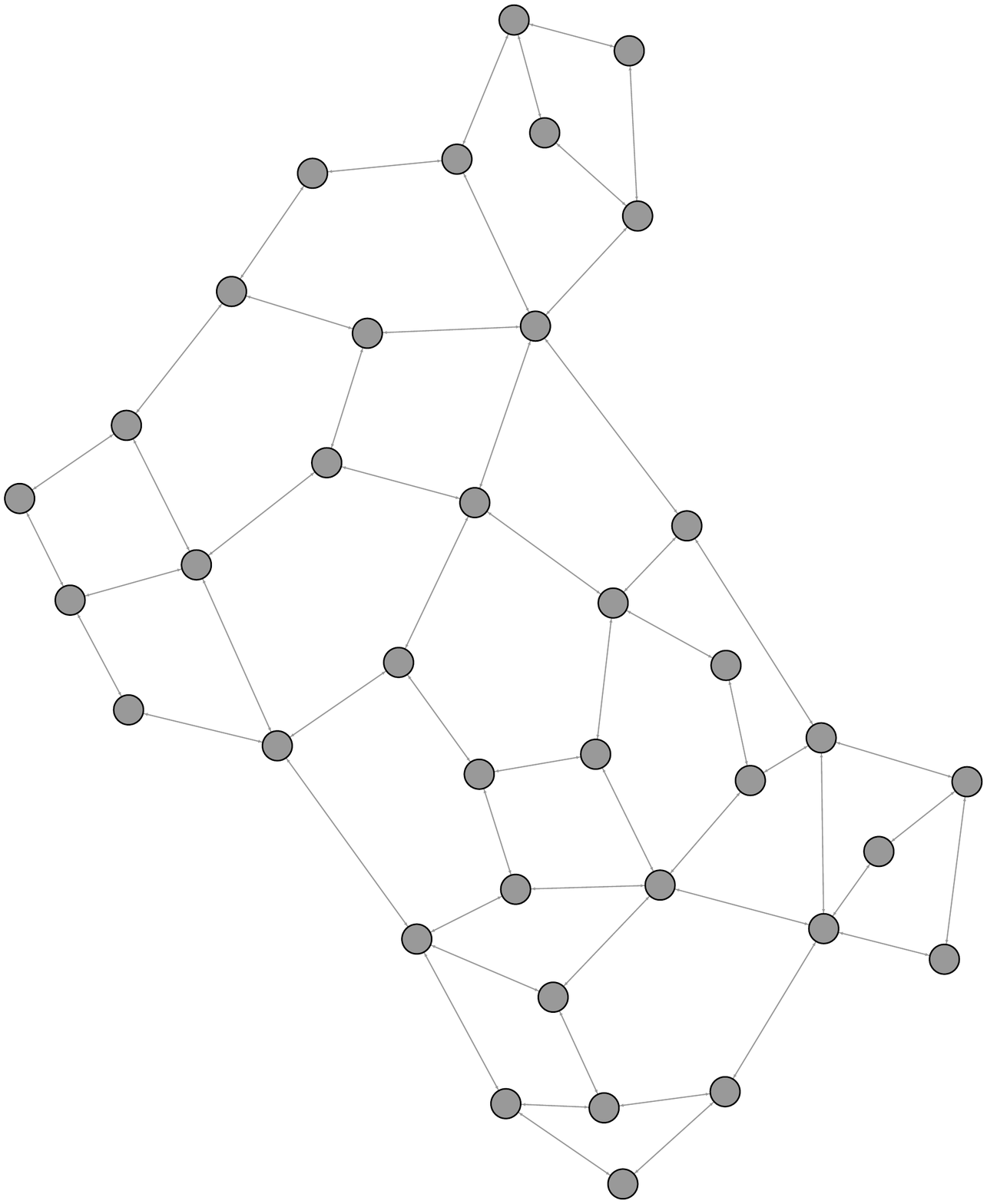}
  \label{fig:cost37}
}
\subfloat[\emph{europg1494}]{
  \includegraphics[width=0.22\textwidth]{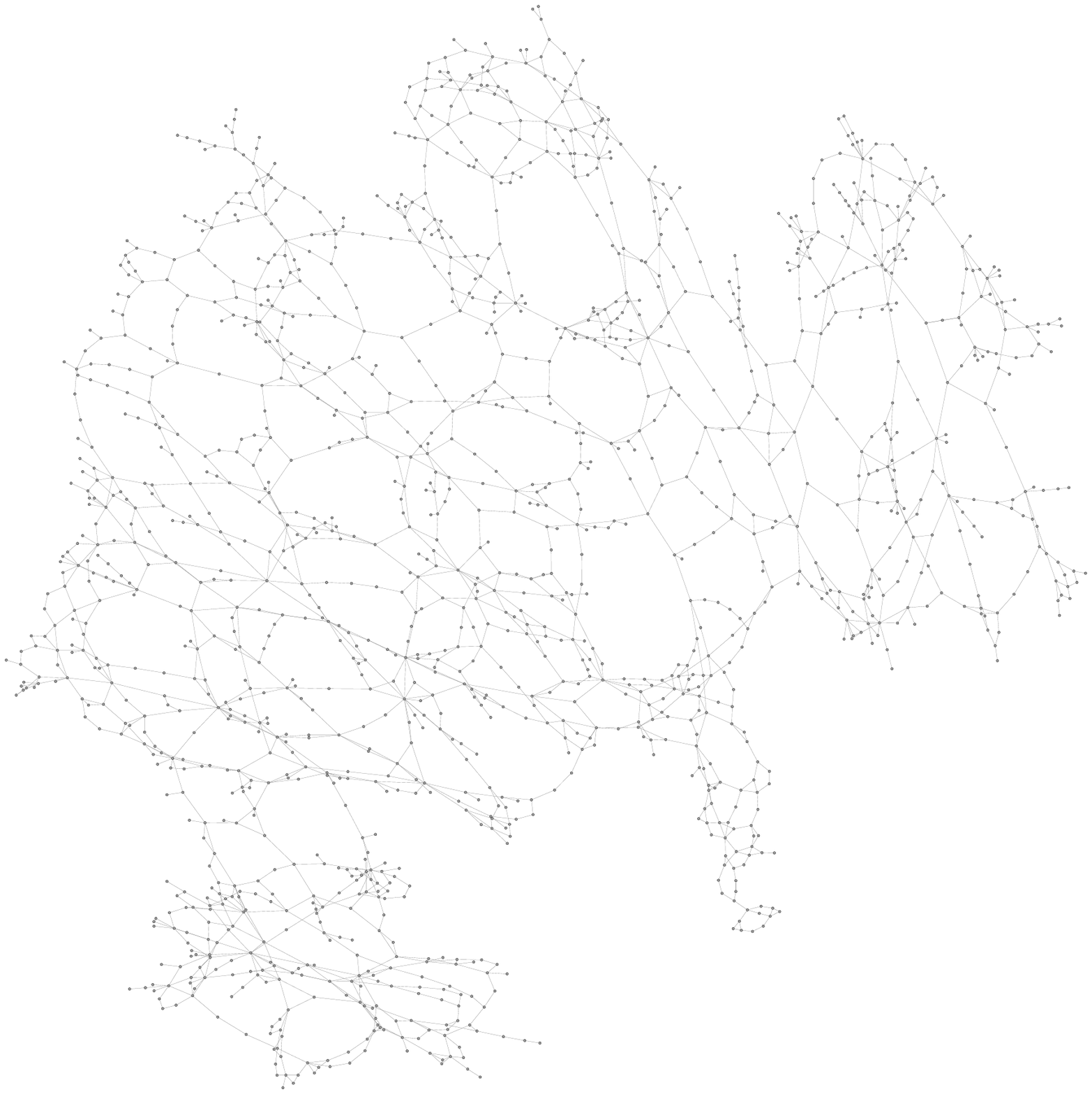}
  \label{fig:europg1494}
}
\subfloat[\emph{fb4039}]{
  \includegraphics[width=0.22\textwidth]{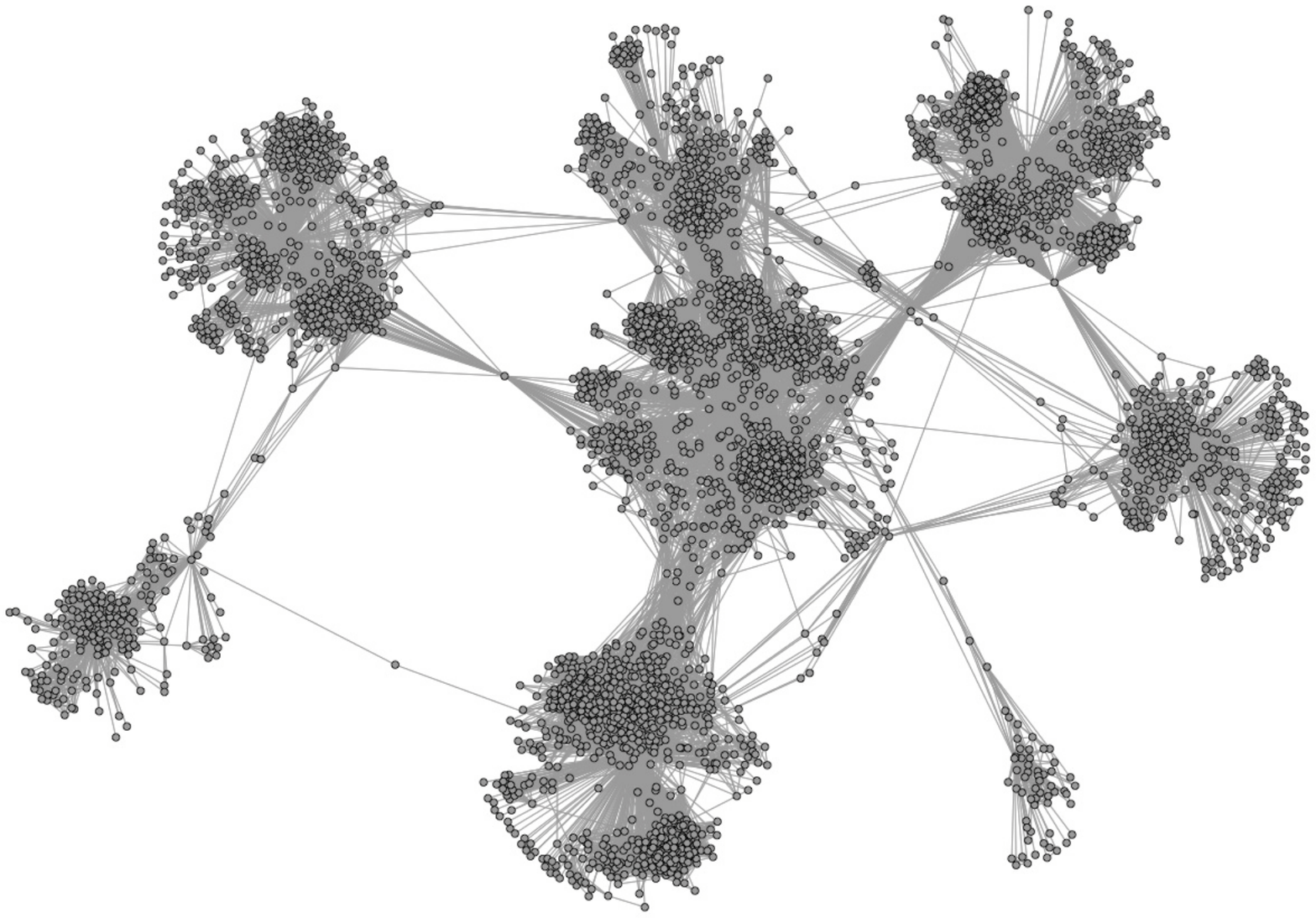}
  \label{fig:fb4039}
}
\\
\subfloat[\emph{wspg4941}]{
  \includegraphics[width=0.22\textwidth]{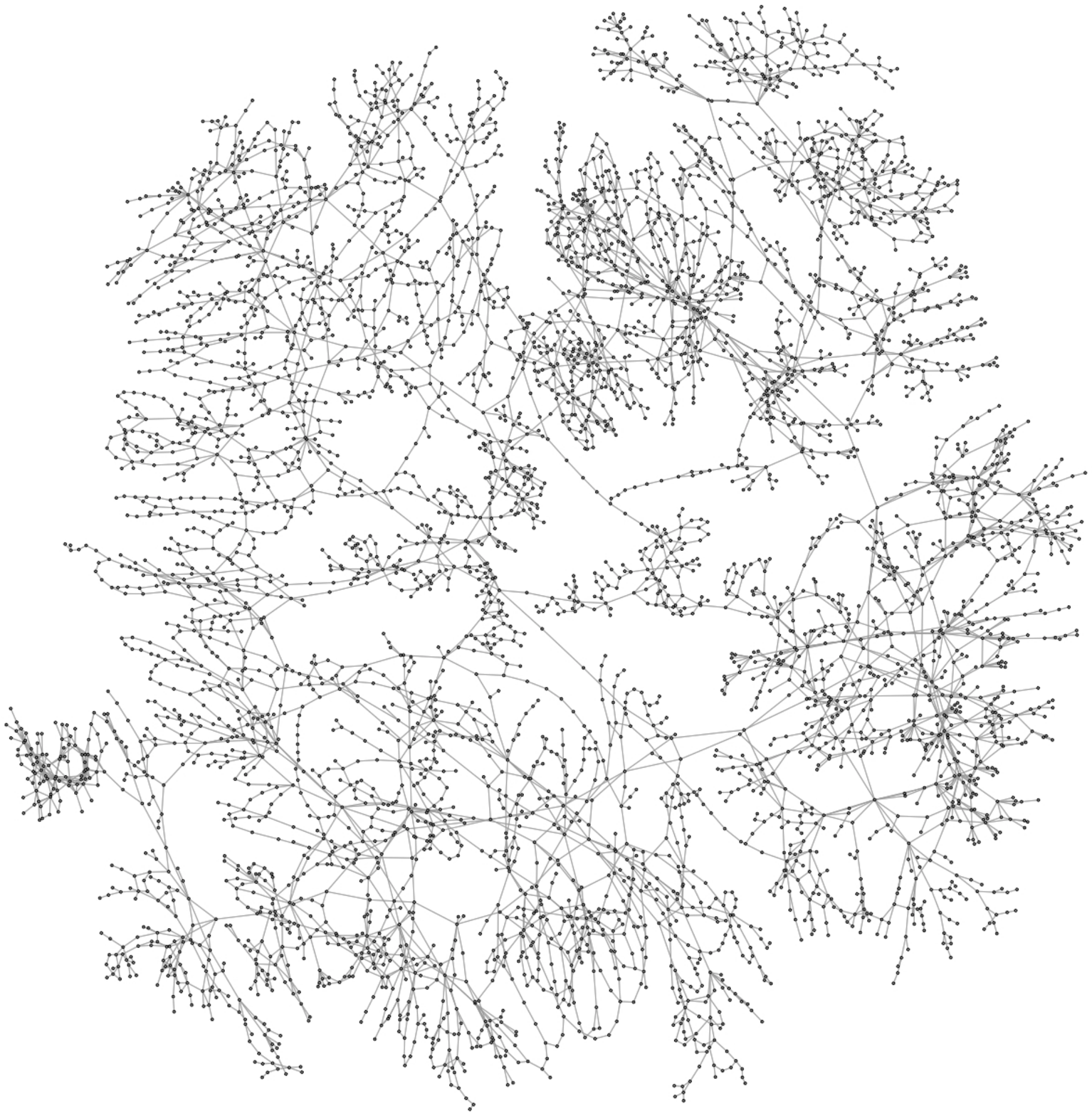}
  \label{fig:wspg4941}
}
\subfloat[\emph{pgieee118}]{
  \includegraphics[width=0.22\textwidth]{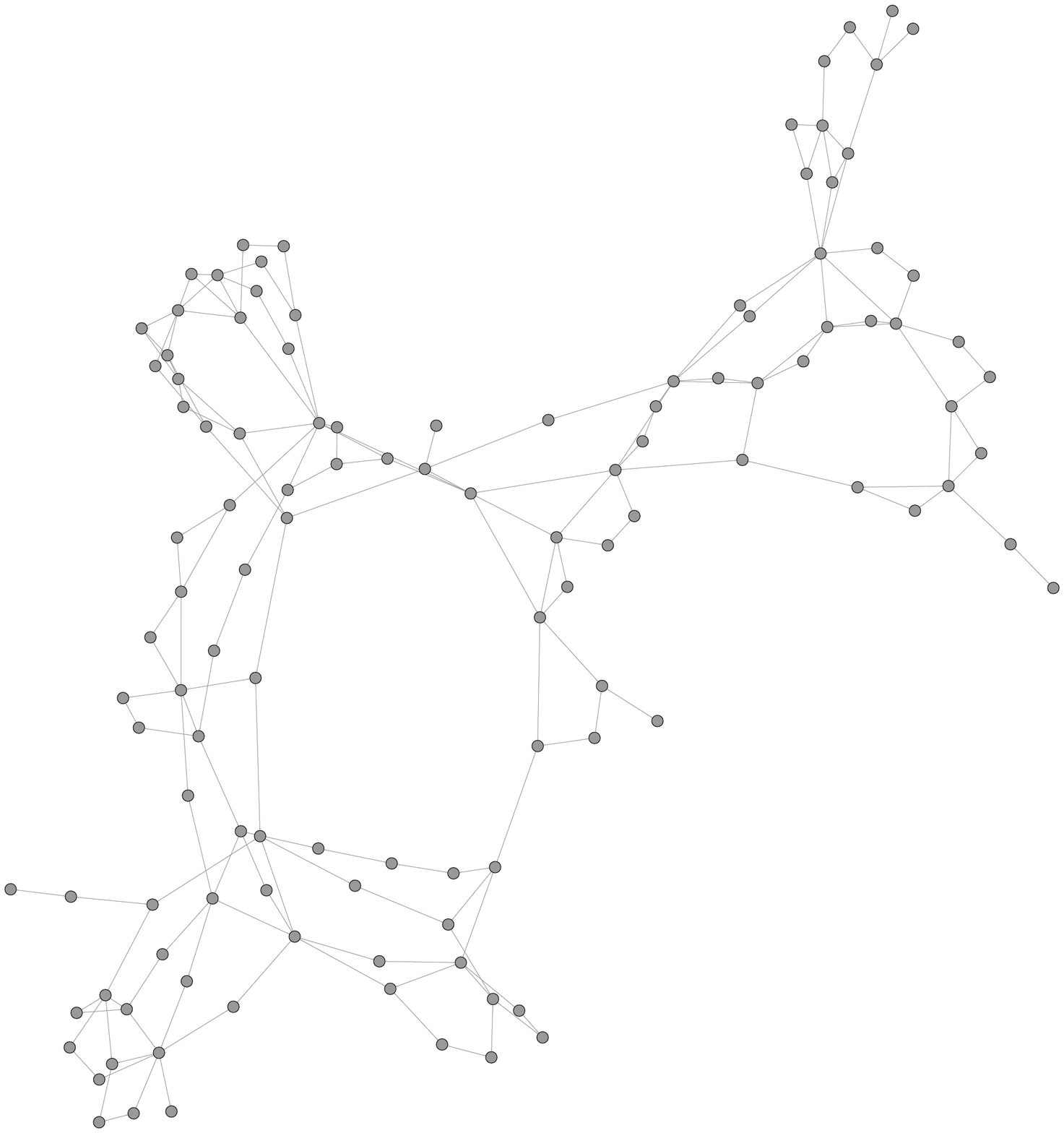}
  \label{fig:pgieee118}
}
\subfloat[\emph{pgieee300}]{
  \includegraphics[width=0.22\textwidth]{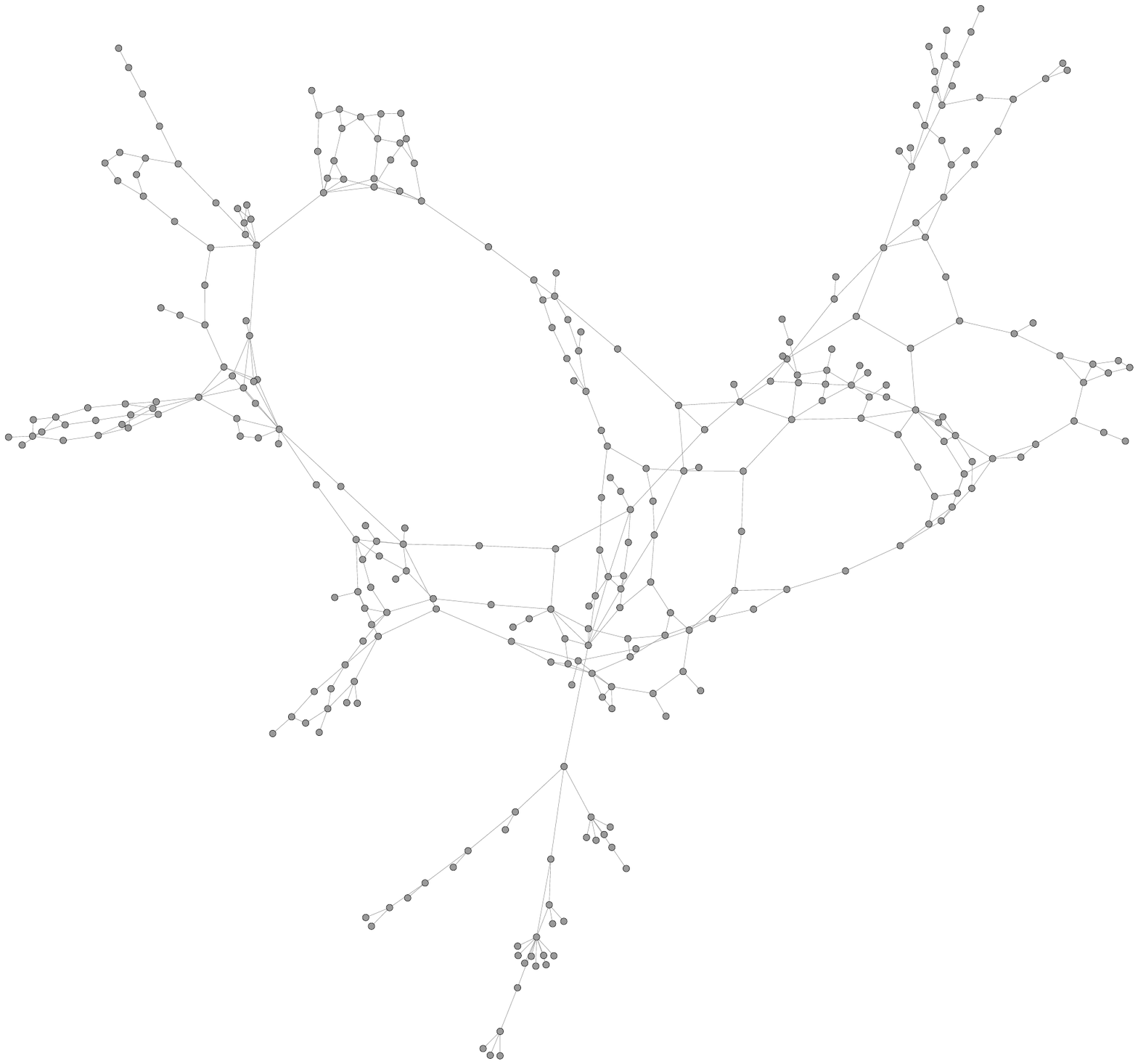}
  \label{fig:pgieee300}
}
  \caption{Layout of 7 out of the 17 topologies considered in this work.}
  \label{ref:top_layout2}
\end{figure}

In this section we present the set of seventeen network topologies considered in our work. These networks have been chosen in order to represent a wide variety of complex network topology types. Generating representative synthetic topologies is a difficult task (and it is not the objective of this paper). Thus, we have conducted an extensive investigation and we have obtained seventeen networks from several sources, which are described next (the name of each network includes the number of nodes):

\begin{enumerate}
	\item \emph{abilene93} (Fig.~\ref{fig:abilene93}): a small network that has been chosen because of its underlying AS topology structure.
	\item \emph{cogentco197} (Fig.~\ref{fig:cogentco197}): a real telecommunications network that has been taken from the repository provided in \cite{topologyzoo}.
	\item \emph{er400} (Fig.~\ref{fig:er400}): a random network that has been generated using the Erd\H{o}s-R\'{e}nyi model \cite{bollobas}.
	\item \emph{powerlaw400} (Fig.~\ref{fig:powerlaw400}): a power-law network that has been generated using the Barab\'{a}si-Albert (BA, preferential attachment mechanism) model \cite{BA}.
	\item \emph{homoge400} (Fig.~\ref{fig:homoge400}): a homogeneous network (a network where all the nodes have equal node degree) that has been generated, being a toroidally-periodic rectangular lattice of size $20 \times 20$. Although this network is not a complex network, it has been considered for comparison purposes.
	\item \emph{bt400} (Fig.~\ref{fig:bt400}): this topology has been obtained by manipulating a previously generated topology using BRITE.
	\item \emph{bo1458} (Fig.~\ref{fig:bo1458}): a protein interaction network for yeast \cite{jeong2001lethality}.	
	\item \emph{col4158} (Fig.~\ref{fig:col4158}): a collaboration network of Arxiv's \emph{General Relativity} category \cite{Leskovec2007}.	
	\item \emph{col8638} (Fig.~\ref{fig:col8638}): a collaboration network of Arxiv's \emph{High Energy Physics Theory} category \cite{Leskovec2007}.	
	\item \emph{cost37} (Fig.~\ref{fig:cost37}): a Pan-european communications reference network.	
	\item \emph{europg1494} (Fig.~\ref{fig:europg1494}): an approximated model of the european power grid network \cite{bialek2013}.	
	\item \emph{fb4039} (Fig.~\ref{fig:fb4039}): this network represents \emph{circles} or \emph{friends list} of the popular social network Facebook \cite{conf/nips/McAuleyL12}.	
	\item \emph{wspg4941} (Fig.~\ref{fig:wspg4941}): a topology of the Western States Power Grid of the United States \cite{Watts1998}.	
	\item \emph{pgieee118} and \emph{pgieee300} (Fig.~\ref{fig:pgieee118} and Fig.~\ref{fig:pgieee300}): these two topologies are reference IEEE power grid networks \cite{referencepg}.	
	\item \emph{AS25357}: an AS network from 2012 \cite{dimesproject}.	
	\item \emph{AS26475}: this network is the largest CAIDA AS connected graph from the network set available in November 2007 \cite{Leskovec2007}.
\end{enumerate}

The \emph{col4158}, \emph{col8638}, \emph{fb4039} and \emph{AS26475} networks have been obtained from the SNAP dataset \cite{SNAP}. The layouts of 15 of the 17 topologies can be observed in Figs.~\ref{ref:top_layout} and \ref{ref:top_layout2}. All of the networks are connected and considered as symmetric graphs. It is worth noting that some of the networks were not connected, and a post-processing has been done in order to obtain the largest connected component. Table~\ref{tab:components} shows the networks that have been post-processed because they were disconnected. Furthermore, Table~\ref{tab:topology_charact1} and Table~\ref{tab:topology_charact2} present several characteristics of this set of networks, some of which are presented with their \emph{standard deviation}. As it can be observed, we have considered a heterogeneous set of networks with respect to the number of nodes, ranging from 37 to 26475.

\begin{table}
	\caption{Networks that were disconnected and for which a post-processing has been done to obtain the largest connected component.\label{tab:components}}
{\begin{center}\begin{tabular}{|l|c|}
  \hline
  Topology & List of $|N| \times $ number of components  \\ \hline\noalign{\smallskip}\hline
\emph{bo1458} & $1458 \times 1$; $7 \times 4$; $6 \times 3$; $5 \times 5$; $4 \times 10$; $3 \times 25$; $2 \times 101$; $1 \times 24$\\
\emph{col4158} & $4158 \times 1$; $14 \times 1$; $12 \times 1$; $10 \times 1$; $9 \times 2$; $8 \times 6$; \\
& $7 \times 8$; $6 \times 12$; $5 \times 17$; $4 \times 30$; $3 \times 98$; $2 \times 177$; $1 \times 1$\\
\emph{col8638} & $8638 \times 1$; $21 \times 1$; $11 \times 1$; $9 \times 2$; $8 \times 6$; $ 7 \times 11$; \\
& $6 \times 8$; $5 \times 21$; $4 \times 45$; $3 \times 67$; $ 2 \times 264$; $1 \times 2$ \\
\emph{europg1494} & $1494 \times 1$; $1 \times 19$ \\
\emph{AS25357} & $25357 \times 1$; $2 \times 5$\\
  \hline
\end{tabular}\end{center}}
\end{table}

The \emph{fb4039} network shows the highest average nodal degree ($\langle k \rangle = 43.69$), what means that every person has an average of about 44 friends in this social network. The two AS networks (\emph{AS25357} and \emph{AS26475}) present the highest mean degree of first neighbors ($\langle d \rangle$) and maximum degree ($k_{\max}$), i.e. in \emph{AS25357} there is an AS that is connected to other ($k_{\max}$) 3781 ASes, and some of them have a high node degree as well. A high $k_{\max}$ is an indicator of vulnerability, depicting that removal of such a node could seriously damage the network. Networks with high values of the largest eigenvalue of the adjacency matrix (or spectral radius, $\lambda_1$) and algebraic connectivity ($\mu_{N-1}$) are more robust. In this case, the \emph{fb4039} network shows the highest spectral radius and the \emph{er400} presents the highest algebraic connectivity. For this reason, these two networks are supposed to be most robust than the rest of them in the case of failures.

\begin{table}
	\caption{Main network features. The table displays, from left to right: topology name, number of nodes, average nodal degree $\pm$
	 \textit{standard deviation} (StDev), mean degree of first neighbors $\pm$ StDev, largest eigenvalue of the adjacency matrix, maximum degree $k_{\max}$ and the second smallest eigenvalue of the Laplacian matrix (the so-called \emph{algebraic connectivity}). 
	\label{tab:topology_charact1}}
{\begin{center}\begin{tabular}{|l|c|c|c|c|c|c|c|c|}
  \hline
  Topology & $N$ & $\langle k \rangle$ $\pm$
StDev & $\langle d \rangle$ $\pm$
StDev & $\lambda_1$ & $k_{\max}$ & $\mu_{N-1}$
  \\ \hline\noalign{\smallskip}\hline
\emph{abilene93}   &   93  &   $2.88\; {\pm}2.71$   & $6.76\; {\pm}2.76$ &   5.016   &  12 & 0.07607 \\
\emph{cogentco197}    &   197 &   $2.46\; {\pm}1.05$   & $2.91\; {\pm}0.92$ &  3.778   &   9 & 0.00858  \\
\emph{er400}   &   400 &   $7.81\; {\pm}2.80$   & $8.89\; {\pm}1.01$ &  8.848   &  15  & 0.90416 \\
\emph{powerlaw400} &   400 &   $2.00\; {\pm}3.25$   & $9.47\; {\pm}11.81$ &  7.013   &   47 & 0.00463 \\
\emph{homoge400}   &   400 &   $4.00\; {\pm}0.00$   & $4.00\; {\pm}0.00$ &  4.000   &   4  & 0.09788\\
\emph{bt400}    &   400    &   $3.74\; {\pm}2.17$   & $5.44\; {\pm}1.61$ &  5.195  & 11 & 0.01013\\
\emph{bo1458} & 1458 & $2.67\; {\pm}3.45$ & $9.65\; {\pm}10.74$ & 7.535 & 56 & 0.02126\\
\emph{col4158} & 4158  & $6.45\; {\pm}8.62$ & $11.60\; {\pm}9.02$ & 45.616 & 81 & 0.03530\\
\emph{col8638} & 8638 & $5.74\; {\pm}6.45$ & $11.25\; {\pm}6.65$ & 31.034 & 65 & 0.02441\\
\emph{cost37} & 37 & $3.08\; {\pm}0.85$ & $3.31\; {\pm}0.45$ & 3.399 & 5 & 0.15857\\
\emph{europg1494} &  1494 & $2.88\; {\pm}1.75$ & $4.17\; {\pm}1.58$ & 5.027 & 13 & 0.00170 \\
\emph{fb4039} & 4039 & $43.69\; {\pm}52.41$ & $105.55\; {\pm}91.30$ & 162.373 & 1045 & 0.01812\\
\emph{wspg4941} &  4941 & $2.66\; {\pm}1.79$ & $3.96\; {\pm}1.93$ & 7.483&19 & 0.00076\\
\emph{pgieee118} & 118 & $3.03\; {\pm}1.56$ & $3.95\; {\pm}1.13$ & 4.105 & 9 & 0.02714 \\
\emph{pgieee300} & 300 & $2.72\; {\pm}1.54$ & $3.86\; {\pm}1.71$ & 4.126 & 11 & 0.00938\\
\emph{AS25357} & 25357 & $5.91\; {\pm}48.03$ & $659.73\; {\pm}827.98$ & 103.361 &3781 & 0.10768\\
\emph{AS26475} & 26475 & $4.03\; {\pm}33.37$ & $471.27\; {\pm}644.72$ & 69.642 & 2628 & 0.02043\\
  \hline
\end{tabular}\end{center}}
\end{table}

\begin{table}
	\caption{Network features. The table displays, from left to right: topology name, average shortest path
	length $\pm$ StDev, normalized average betweenness centrality $\pm$ StDev, average clustering
	coefficient $\pm$ StDev, and assortativity coefficient $|r|\leq 1$.\label{tab:topology_charact2}}
{\begin{center}\begin{tabular}{|l|r|c|c|c|}
  \hline
  Topology & $\langle l \rangle$ $\pm$ StDev & $\langle b \rangle$ $\pm$ StDev & $\langle C \rangle$ $\pm$ StDev & $r$
  \\ \hline\noalign{\smallskip}\hline
\emph{abilene93}   &   $   3.92 \;   {\pm}   1.32    $   &   $   0.0529 \; {\pm}   0.0551  $   &   $   0.51 \;  {\pm}   0.48   $   &   $-0.5130$ \\
\emph{cogentco197}    &   $   10.52 \;  {\pm}   5.09    $   &   $   0.0585 \; {\pm}   0.0665  $   &   $   0.12\;   {\pm}   0.32   $   &   $+0.01956$  \\
\emph{er400}   &   $   3.13  \;  {\pm}   0.73    $   &   $   0.0103 \; {\pm}   0.0037  $   &   $   0.02\;   {\pm}   0.07   $   &   $-0.07229$ \\
\emph{powerlaw400} &   $   6.01  \;  {\pm}   2.16    $   &   $   0.0175 \; {\pm}   0.0594  $   &   $   0.64 \;  {\pm}   0.47   $   &   $-0.16512$ \\
\emph{homoge400}   &   $   10.03  \; {\pm}   4.10    $   &   $   0.0276  \;{\pm}   0.0000  $   &   $   0.00 \;  {\pm}   0.00   $   &   $+1.0000$  \\
\emph{bt400}    &   $   10.12  \;  {\pm}   4.21    $   &   $   0.0202 \; {\pm}   0.0357  $   &   $   0.16 \;  {\pm}   0.27   $   &   $-0.29646$ \\
\emph{bo1458} & $ 6.81\;  {\pm} 2.04 $ & $ 0.0039\;  {\pm} 0.0110 $ & $ 0.56\;  {\pm} 0.47 $ & $-0.20954$\\
\emph{col4158} & $ 6.04\;  {\pm} 1.57 $ & $ 0.0012 \;  {\pm} 0.0034 $ & $ 0.71\;  {\pm} 0.35 $ & $+0.63919$\\
\emph{col8638} & $5.94 \;  {\pm} 1.50 $ & $ 0.0005\;  {\pm} 0.0015 $ & $ 0.65\;  {\pm}0.37  $ & $ +0.23892	$\\
\emph{cost37} & $ 4.05\;  {\pm} 1.90 $ & $0.0782 \;  {\pm} 0.0756 $ & $ 0.00\;  {\pm} 0.00 $ & $ -0.01510	$\\
\emph{europg1494} & $18.88 \;  {\pm} 8.73 $ & $ 0.0119\;  {\pm}  0.0304$ & $ 0.27\;  {\pm} 0.40 $ & $ -0.11965$\\
\emph{fb4039} & $ 3.69\;  {\pm} 1.19 $ & $0.0006 \;  {\pm} 0.0116 $ & $ 0.62\;  {\pm} 0.20 $ & $ +0.06358$\\
\emph{wspg4941} & $ 18.98\;  {\pm} 6.50$ & $0.0036 \;  {\pm} 0.0160 $ & $ 0.32\;  {\pm} 0.44 $ & $ +0.00346$\\
\emph{pgieee118} & $6.30\;  {\pm}2.81  $ & $0.0457\;  {\pm} 0.0723 $ & $0.22 \;  {\pm} 0.36 $ & $-0.15257$\\
\emph{pgieee300} & $ 9.93\;  {\pm} 4.06 $ & $ 0.0299\;  {\pm} 0.0546 $ & $ 0.31\;  {\pm}  0.42$ & $-0.22063 $\\
\emph{AS25357} & $3.39 \;  {\pm} 0.70 $ & $ 0.0001\;  {\pm} 0.0020 $ & $0.73 \;  {\pm} 0.36 $ & $ -0.18540$\\
\emph{AS26475} & $3.87 \;  {\pm}0.90  $ & $ 0.0001\;  {\pm}0.0020  $ & $ 0.58\;  {\pm} 0.46 $ & $-0.19465 $\\
  \hline
\end{tabular}\end{center}}
\end{table}

Regarding the average shortest-path length ($\langle l \rangle$) it is shown that two power grid networks (\emph{europg1494} and \emph{wspg4941}) have the higher values and consequently are more vulnerable. This is due to the fact that, traditionally, power grid networks have a tree-like structure. Furthermore, the average node betweenness centrality ($\langle b \rangle$) of \emph{cost37}, \emph{cogentco197} and \emph{abilene93} shows that these three topologies have an excess of centrality measures for some nodes, indicating the vulnerability of networks under targeted failures. The absence of 3-cycles in the clustering coefficient ($\langle C \rangle$) measurements reveal that the \emph{homoge400} and \emph{cost37} lack two-hop paths to re-route the traffic in case of failure of one of its neighbors. Finally, networks with negative values of assortativity ($r$) have an excess of radial links, i.e., links connecting nodes of dissimilar degrees. Such a property is typical of technological networks \cite{newman2003structure}. 

This initial network analysis of the considered set of topologies reveals that none of the networks can be considered as the most robust for all of the metrics. Besides, the vulnerability of the networks is going to differ depending on the considered type of multiple failures. As a consequence, it is necessary to define new metrics able to characterize how robust a network is in a specific scenario. The following two sections present two new measures to evaluate network vulnerability in the case of epidemic-like and cascading failures.

\section{Epidemic-like failures\label{sec:epidemics}}

Throughout the history of mankind there have been many diseases that have spread quickly, becoming an epidemic or even a pandemic. As a result, many epidemic outbreaks have ravaged human civilizations from the Middle Ages until today. For instance, the devastating Influenza epidemic of 1918 (the third greatest plague in history) claimed 21 million lives and affected over half the world's population \cite{marks1976epidemics}.

Epidemic models are used to model the spreading of events (e.g. failures) in several types of complex networks. These models have been used in a wide variety of research fields. For instance, in \cite{pawel2013forestfire} the authors used characteristics of epidemic spreading to model the fire propagation on a forest. In \cite{hill2010social}, the authors used epidemic models to show that emotional states spread like infectious diseases across social networks. In \cite{Montanari2010spreading} it was shown that there are certain network structures that facilitate the propagation of new ideas, behaviors or technologies. In the last years, online social networks (OSNs) have also been the focus of study. For instance, in \cite{Liu2013virus} the authors studied how to control virus propagation in OSNs. Finally, although no commercial references (or reports) have been found with respect to the propagation of failures in telecommunication networks, several works have focused on analyzing the consequences of epidemic attacks on the services provided by such networks \cite{calle2010multiple}, \cite{manzano2011quantitative}, \cite{manzano2012vilanova}. Additionally, a framework to eradicate epidemic failure has been recently proposed in \cite{manzano2012epidemic}. Nonetheless, to the best of our knowledge, no methods to detect the most vulnerable nodes of a complex network in the case of epidemic failures have been proposed. Therefore, a first step would be to define network measures to characterize all nodes under such failure scenarios.

\subsection{Epidemic Models\label{sec:relatedwork}}

Epidemic dynamics in complex networks have undergone extensive research \cite{PS2001epidemic}, \cite{Newman2002spread} \cite{PS2002epidemic}, \cite{Bogu2002epidemic}, \cite{epidthresh}. As a consequence, many epidemic models have been proposed and several families are described in the literature (see Chapter~8 in \cite{lewis2010netscience}, Chapter~17 in \cite{Newman2010} and Chapter~14 in \cite{cohen2010complex}). The first family, called \emph{Susceptible-Infected} (SI) considers individuals as being either susceptible (S) or infected (I). This family assumes that the infected individuals will remain infected forever, and so can be used for worst case propagation ($S\rightarrow I$). Another family is the \textit{Susceptible-Infected-Susceptible} (SIS) group, which considers that a susceptible individual can become infected on contact with another infected individual, then recovers with some probability of becoming susceptible again. Therefore, individuals will change their state from susceptible to infected, and vice versa, several times ($S\leftrightarrows I$). The \textit{Susceptible-Exposed-Infected-Susceptible} (SEIS) model is based on the SIS model, and takes into consideration the exposed or latent period of the disease ($S\rightarrow E \rightarrow I \rightarrow S$). The third broad family is \textit{Susceptible-Infected-Removed} (SIR), which extends the SI model to take into account a removed state. In the SIR model, an individual can be infected just once because when the infected individual recovers, becomes either immune or dead, and will no longer pass the infection onto others ($S\rightarrow I \rightarrow R$). Finally, there are two families that extend the SIR family: \emph{Susceptible-Infected-Detected-Removed} (SIDR) and \emph{Susceptible-Infected-Removed-Susceptible} (SIRS). The first one adds a Detected (D) state, and is used to study virus throttling, which is an automatic mechanism for restraining or slowing down the spread of diseases ($S\rightarrow I \rightarrow D \rightarrow R$). The second one considers that after an individual becomes removed, it remains in that state for a specific period of time and then goes back to the susceptible state ($S\rightarrow I \rightarrow R \rightarrow S$). 

Regarding communication networks, an extension of the SIS model, which is called Susceptible-Infected-Disabled-Susceptible (SIDS), was proposed in \cite{calle2010multiple} in order to overcome the limitations of the SIS model with respect to optical transport networks. The SIDS model (\textit{Susceptible$\leftrightarrows$Infected$\rightarrow$Disabled$\rightarrow$Susceptible}) is proposed as one of the first models to consider real telecommunication networks features and it relates each state to a functionality of the network devices. In addition, other epidemic models have also been proposed for wireless telecommunication networks \cite{Myoung2010}.

In this paper we propose a new network measure taking into account the SIS model, which is characterized by two probabilities: 
\begin{inparaenum}[(a)]
	\item $\beta$, the probability of being infected by an already infected node; and 
	\item $\delta$, the probability of an infected node to recover and become susceptible again. 
\end{inparaenum} However, our proposal can be also applied to any other epidemic model and we plan to do so in the future.

Furthermore, according to \cite{epidthresh} and from the following equation:

\begin{equation}
s=\frac{\beta}{\delta} \lambda_1
\end{equation}
where $s$ is the \emph{epidemic intensity} and $\lambda_1$ is the network's largest eigenvalue of the adjacency matrix, which has been typically used to predict network robustness, when $s > 1$ an epidemic survives and the spread of the infection might never die. Thus, in order to obtain comparable results between networks with respect to our proposal (\emph{epidemic survivability}), $s$ must be a parameter of our new measure.

In this work we have fixed $s=3$ for all networks, in order to obtain comparable results, and we have obtained a specific $\beta$ value for each network from the equation $\beta=\frac{s\delta}{\lambda}$.

\subsection{Epidemic Survivability\label{sec:metric}}

Here we present our new network measure called \emph{epidemic survivability} ($ES$). We define our proposal as the probability for each node of a given network to be eventually infected (i.e., in a large enough amount of time steps), given a specific \emph{epidemic intensity} ($s$). This probability of each node asymptotically reaches a stationary state, according to simulations and theoretical models. \emph{Epidemic survivability} can be described as the proportion of time for which each node of a given network has been infected for a given $s$, in a large enough period of time, as shown in Eq.~\ref{ref:eq_ES}:

\begin{equation}\label{ref:eq_ES}
	ES_{i}(s) = \frac{\mbox{time for which node } i \mbox{ has been infected}}{\mbox{total time}}\, \qquad i=1, \dots , N 
\end{equation}
where $N$ is the number of nodes of the network. As a result, $ES$ has a value between 0 and 1 for each node, where higher the value, more vulnerable is the node under the specified epidemic scenario. Formally, from the SIS model, \emph{epidemic survivability} can be computed with the following equation:
\begin{equation} \label{ref:eq_form}
	ES^{*}_{i} = \frac{1}{1+(\frac{\beta}{\delta}\sum_{j\sim i} ES^{*}_{j})^{-1}}\, \qquad i=1, \dots , N 
\end{equation}
where $*$ means \emph{at the stationary state} and $j \sim i $ is the set of \emph{neighbors} of node $i$. Here, it is assumed that $\delta$ and $s$ are given as parameters and $\beta$ is obtained from the equation $\beta=\frac{s\delta}{\lambda_1}$. Thus, it can be observed that Eq.~\ref{ref:eq_form} is a recursive formula and must be initialized with a value. We define this initialization of the probabilities in Eq.~\ref{ref:init}:

\begin{equation}\label{ref:init}
ES^*_{i,\textrm{approx}} = (1- \frac{1}{s}) \, \qquad i=1, \dots , N 
\end{equation}
which corresponds to the solution of Eq.~\ref{ref:eq_form} for the case of a homogeneous/regular network. Moreover, a procedure for computing \emph{epidemic survivability} is provided in Algorithm~\ref{alg}. As it can be observed, the method requires five parameters: the network $\mathcal{G}$ and four constants ($s$, $\delta$, $k$ and $tol$). The first two steps (lines~\ref{l1} and \ref{l2}) compute the largest eigenvalue of the given network and thus obtain the $\beta$ value of the epidemic model. Then, all probabilities are initialized as stated in Eq.~\ref{ref:init} (lines~\ref{l3} to \ref{l4}). Therefore, in the main loop of line~\ref{l5} the new probability of each node is computed as defined in Eq.~\ref{ref:eq_form} (lines~\ref{l6} to \ref{l7}). After that, the absolute error is checked (lines~\ref{l8} to \ref{l9}) and if it results lower than the given tolerance ($tol$) then the algorithm ends, and returns the array containing the \emph{epidemic survivability} of each node of the network. If the absolute error is still higher than $tol$ another iteration is performed.

\algsetup{indent=2em}
\begin{algorithm}
	\caption{Compute \emph{epidemic survivability}.}
	\label{alg}
	\begin{algorithmic}[1]
		\REQUIRE $s \geq 1, d > 0, k > 0, tol > 0, connected~\mathcal{G}$
		\STATE \textnormal{\textbf{Input:} a graph $\mathcal{G}$ and the constants $s$ (epidemic intensity), $\delta$ (repairing rate), $k$ (maximum number of iterations) and $tol$ (tolerance).}
		\STATE \textnormal{\textbf{Output:} an array containing the \emph{epidemic survivability} of each node.}

		\STATE $\lambda\gets$ spectralRadius($\mathcal{G}$) \COMMENT{largest eigenvalue} \label{l1}
		\STATE $\beta\gets$ $\frac{s*\delta}{\lambda}$ \label{l2}
			
		\FORALL{$v \in vertexSet(\mathcal{G})$} \label{l3}
		    \STATE $ P_{ES}[v] = (1- \frac{1}{s})$
		 \ENDFOR \label{l4}
		
	 	\FOR{$c=1 \to k$} \label{l5}
			\FORALL{$v \in vertexSet(\mathcal{G}) $} \label{l6}
			    \STATE $P_{aux}[v] = \frac{1}{1+(\frac{\beta}{\delta}\sum_{j\sim v} P_{ES}[j])^{-1}}$
			 \ENDFOR \label{l7}
			\IF{$(\|P_{aux} - P_{ES}\|) < tol$} \label{l8}
				 \STATE $\mathsf{break}$
			\ENDIF \label{l9}
			
	  		\STATE $P_{ES} \gets P_{aux}$ \label{l10}
	  	\ENDFOR
	\RETURN $P_{ES}$
	\end{algorithmic}
\end{algorithm}

\begin{figure}
\begin{center}
\includegraphics[scale=1.17]{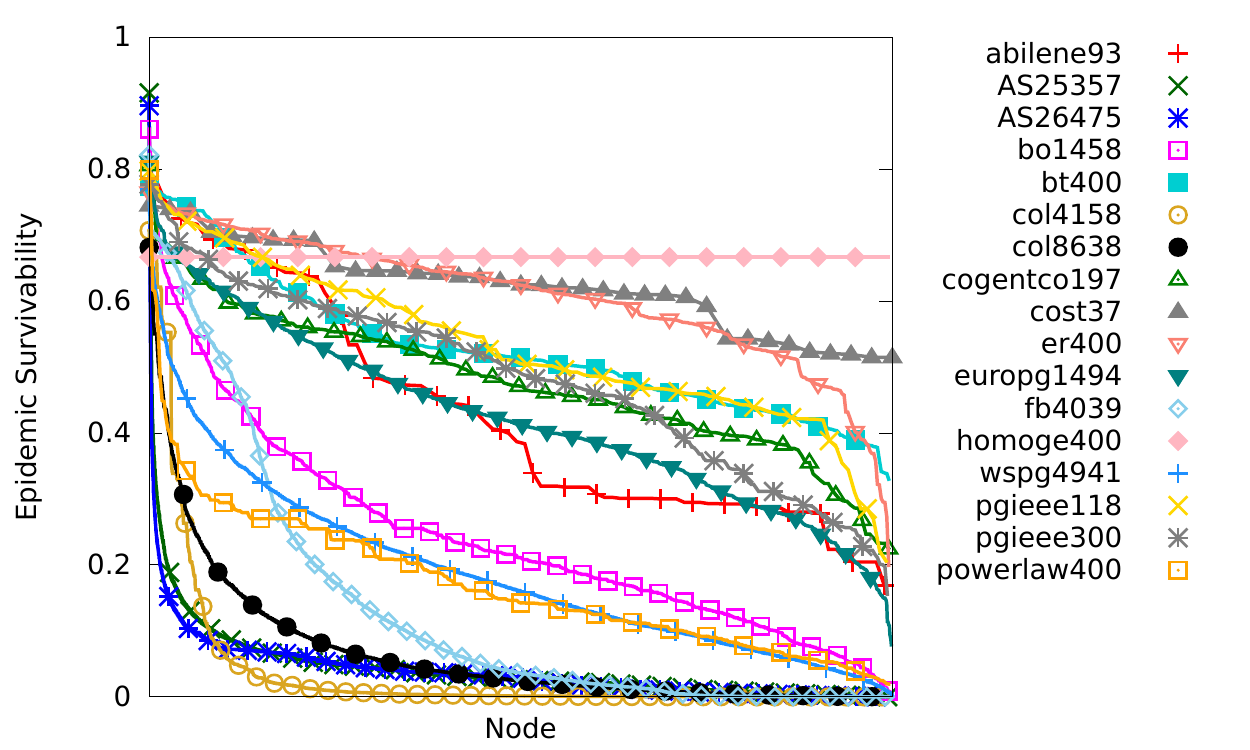}
\caption{\emph{Epidemic survivability} distribution, sorted from major to minor values, of all networks. In this case the set of parameters has been: $s=3$, $\delta=0.3$, $k=2000$ and $tol=1e^{-8}$. The X-axis shows the nodes of the network, their index not showed for the sake of clarity.}
\label{fig:ES_s5}
\end{center}
\end{figure}

\subsection{The distribution\label{dist}}

When computing the \emph{epidemic survivability} for the nodes of a network, according to a specified set of parameters, it is interesting to analyze the distribution of $ES$ values. If these values are sorted, for example, in descending order, it facilitates the comparison between network topologies when considering the same failure propagation scenario for all of them. This approach is illustrated in Fig.~\ref{fig:ES_s5} which displays the \emph{epidemic survivability} distribution, the $ES$ of each node, for the 17 networks in a specific epidemic scenario. As can be observed, the two AS networks (\emph{AS25357} and \emph{AS26475}) together with the two collaboration networks (\emph{col4158} and \emph{col8638}) show the lowest $ES$ distributions, demonstrating that such networks are more robust than the rest of networks, in the case of an epidemic-like failure with epidemic intensity $s=3$. 

It is worth noting that different types of complex networks show different $ES$ distribution curves. While AS and collaboration networks show power-law-like curves, power grids, telecommunication networks, synthetic networks and the biological network depict more smooth-decreasing curves. On one hand, curves showing a rapid decrease (i.e. power-law-like profile) would be expected in complex networks regarding critical infrastructures. This is due to the fact that only a small portion of the nodes of the network would be highly vulnerable, and consequently, it would require less effort (e.g. economical) from the network engineer or operator to protect it. On the other hand, regarding social networks one could expect different curve profiles, depending on the purpose of the social network (e.g. a country's government interested in controlling its social networks would prefer flatter curves, because there would not be any node with a high spreading potential).

\section{Cascading failures\label{sec:cascading}}

A cascading failure event is typically triggered by a single point of failure (i.e. one component) that leads to a domino effect, causing other parts of the network to fail. When such failures occur they can affect significant percentages of the world's population. For instance, according to \cite{cascadeindia}, in 2012 a cascading failure in North India left more than 300 million people without power.

Cascading failure models have been used to understand these dynamic multiple failures in different types of complex networks. The power grids are the most remarkable example where cascading failures can occur. There are several works which have studied the impact of cascading failures on different power grids: Italy \cite{crucitti2004powergrid}, North America \cite{albert04} and Europe \cite{Rosas-Casals2007}. However, cascading failures are not limited to power grids, but any load/capacity related complex network. For example, the authors of \cite{Habib2013cascade} stated that two types of cascading failures can occur in backbone telecommunication networks. Other works such as \cite{Cal2002} and \cite{wrap32818} have focused on the IP layer  and optical layer of communication networks, respectively. Moreover, cascading failures have been also studied in socio-technological networks \cite{10.1371/journal.pone.0045406}. Other examples of cascading failures include biological, electronic and financial networks. 

Although the authors of \cite{Koc:373} proposed a robustness metric for power grid networks in the case of targeted attacks, to the best of our knowledge, there is not any metric which can be generally applied to any kind of cascading failure or complex network. Therefore, with the purpose of providing the network scientific community with such a measure, in this section we define \emph{cascading survivability}.

\subsection{Cascading Failure Models}
Cascading failures have been extensively studied in the literature. Some of the most well-known models are presented next. In \cite{Watts30042002} one of the first cascading failure models was presented, which focused on random complex networks. Contemporarily, the authors of  \cite{motter2002cascading} presented a simple but functional model. Later on, the model was enhanced in \cite{crucitti2004cascading} by keeping an auxiliary cost matrix related with the efficiency metric \cite{Lattora2001efficient,Lai2004}. Furthermore, in \cite{PES:262284} the authors proposed an analytically tractable loading-dependent cascading failure model. In \cite{Nedic2006627} an AC blackout model representing most of the interactions observed in cascading failures was presented. Recently, in \cite{Guo2012cascading} a cascading failure model for inter-domain routing systems was presented. Moreover, the authors proposed two metrics to assess the impact of a cascading failure: the proportion of failure nodes and the proportion of failed links. 

As previously stated in this work, our objective is to define a metric able to characterize the vulnerability of the elements of a network (i.e. in this case nodes) under cascading failures. To do so, we have chosen the model presented in \cite{motter2002cascading}. According to this model, each node $j$ is related with a load $L_j$. The load at each node is the node betweenness centrality, i.e. the number of shortest paths passing through the node. Then, the capacity can be defined as a proportional value to the initial load $L_j$, as denoted by Eq.~\ref{ref:cap}:

\begin{equation}\label{ref:cap}
C_j = (1 + \alpha) \cdot L_j\, \qquad j=1,2,\dots,N
\end{equation}
where $N$ is the number of nodes of the network and $\alpha$, the \emph{tolerance} parameter of the model, is a constant that must be $\alpha \ge 0$. This parameter is related with the concept of \emph{capacity dimensioning} of a network, which is of paramount importance at the designing phase of a network (e.g. a critical infrastructure such as a power grid). An appropriate level of \emph{over-dimensioning} can prevent a network from cascading failures. However, a higher $\alpha$ typically involves a higher economical budget. Therefore, network engineers must seek a trade-off between these two factors.

As defined by the model in \cite{motter2002cascading}, we focus on cascades triggered by the removal of a single node. This event, in general, causes changes in the distribution of shortest paths. As a result, after an initial node failure, the new load of the nodes ($L_j'$) might be different from the initial load ($L_j$). Then, for each node, if the expression of Eq.~\ref{expression} is satisfied:
\begin{equation}\label{expression}
 L_j' > C_j   
\end{equation}
the node $j$ overloads and fails, which might cause subsequent overloading failures on the rest of nodes of the network.

Finally, we note that in the results presented further in this section we have assumed an $\alpha=0.05$ for all networks, with the purpose of allowing comparison among them.

\subsection{Cascading Survivability}
Our new network measure called \emph{cascading survivability} ($CS$) is presented below. \emph{Cascading survivability} evaluates how potentially injurious a node is according to a specific cascading failure scenario. In other words, $CS$ can be described as shown in Eq.~\ref{ref:cs}:

\begin{equation}\label{ref:cs}
CS_{i}(\alpha) = \frac{\textrm{the number of nodes that fail if node } i \textrm{ initially fails}}{\textrm{all nodes in the network} - 1} \, \qquad 1=1, \dots , N 
\end{equation}
where $N$ is the number of nodes of the network. As observed, $\alpha$ is a parameter of $CS$, what means that for different $\alpha$ distinct $CS$ values might be obtained. \emph{Cascading survivability} takes values in the range between 0 and 1 for each node, where higher the value, more harmful is the node under a specific cascading failure scenario. 
\algsetup{indent=2em}
\begin{algorithm}
	\caption{Compute \emph{cascading survivability}.}
	\label{alg1}
	\begin{algorithmic}[1]
		\REQUIRE $\alpha \geq 0, connected~\mathcal{G}$
		\STATE \textnormal{\textbf{Input:} a graph $\mathcal{G}$ and the constant $\alpha$ (tolerance parameter).}
		\STATE \textnormal{\textbf{Output:} an array containing the \emph{cascading survivability} of each node.}
		
		\STATE $N\gets$ vertexSize($\mathcal{G})$)
		
		\COMMENT{initializing load and capacity of each node}
		\FORALL{$v \in vertexSet(\mathcal{G})$} \label{c1} 
		    \STATE $ L[v] = $ NodeBetweennessCentrality($\mathcal{G}$)
			\STATE $C[v] = (1 + \alpha) \cdot L[v]$
		\ENDFOR \label{c2}

		\FORALL{$v \in vertexSet(\mathcal{G})$} \label{c3}
			\STATE $F\gets$ add($v$) \COMMENT{add node $v$ to the list of nodes that are going to fail} \label{c4}
		
			\WHILE{$F$ is not empty } \label{c5}
		    	\STATE $\mathcal{G}'\gets$ removeNodes($\mathcal{G}$,$F$) \COMMENT{removes from $\mathcal{G}$ all nodes in $F$. After the operation $F$ is empty.} \label{c6}

				\FORALL{$u \in vertexSet(\mathcal{G}')$} \label{c7}
			    	\STATE $ L'[u] = $ NodeBetweennessCentrality($\mathcal{G}'$) \label{c8}
			
					\IF{$L'[u] > C[u]$} \label{c9}
					 	\STATE $F\gets$ add($u$) \label{c10}
						\STATE $CS[v]= CS[v] + 1$ \COMMENT{increase in 1 the number of nodes that have failed due to the initial failure of $v$} \label{c11}
					\ENDIF \label{c12}
				\ENDFOR \label{c13}
			\ENDWHILE \label{c14}
 		\ENDFOR \label{c15}

		\FORALL{$v \in vertexSet(\mathcal{G})$} \label{c16}
		    \STATE $CS[v] = \frac{CS[v]}{N-1}$
		\ENDFOR \label{c17}
		
	\RETURN $CS$
	\end{algorithmic}
\end{algorithm}

\begin{figure}
\begin{center}
\includegraphics[scale=1.17]{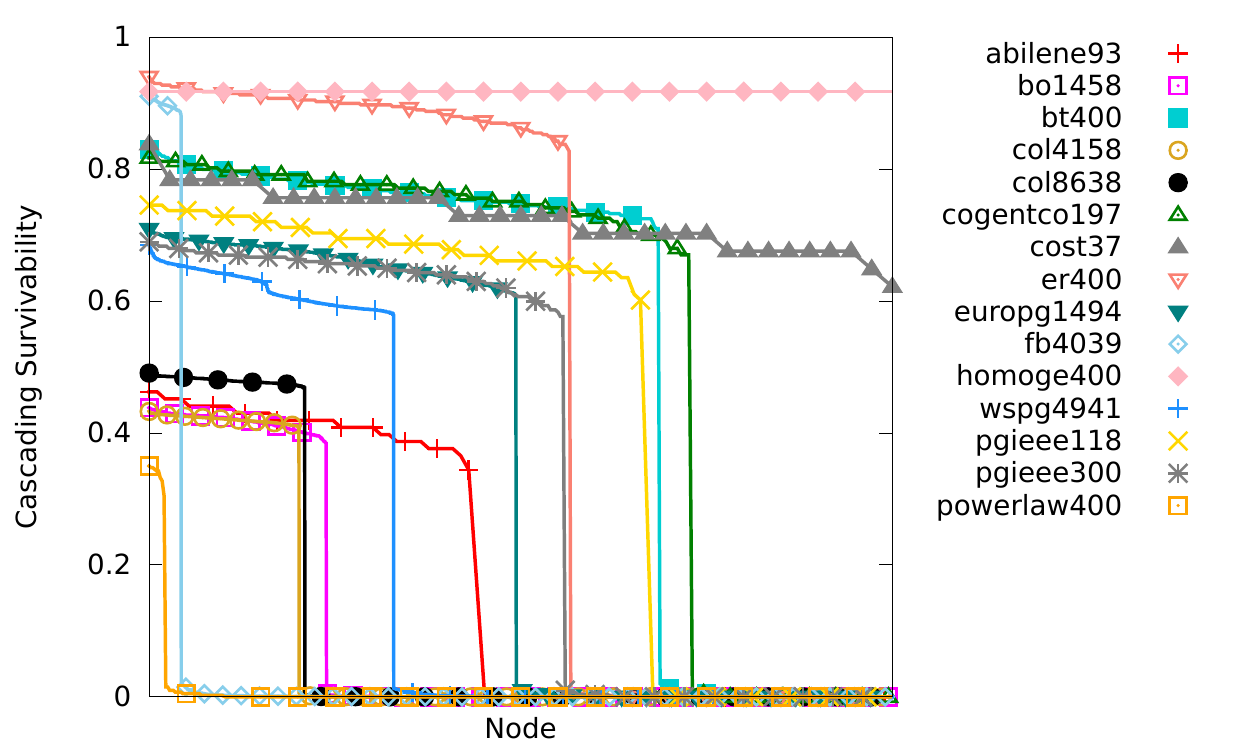}
\caption{\emph{Cascading survivability} distribution, sorted from major to minor values, of all networks. In this case we have considered $alpha=0.05$. The X-axis shows the nodes of the network, their index not showed for the sake of clarity.}
\label{fig:CS}
\end{center}
\end{figure}

We have defined a procedure to compute the \emph{cascading survivability} of the nodes of a network, which is presented in Algorithm~\ref{alg1}. As shown, the method requires two parameters: the network $\mathcal{G}$ and the tolerance parameter $\alpha$. First of all, the initial load and capacity of each node is computed (lines~\ref{c1} to \ref{c2}). Then, an initial failure is caused, for each one of the nodes of the given network, one at a time (lines~\ref{c3} to \ref{c15}). For each initial failure (line~\ref{c4}) and as well as at each step of the spreading of the cascade, (lines~\ref{c5} to \ref{c14}), the new load of the remaining nodes of the network is computed (line~\ref{c8}). If the new load becomes higher than the capacity at any step, then the \emph{cascading survivability} of the node that initially triggered the failure is increased (lines~\ref{c9} to \ref{c12}). Finally, the $CS$ of each node is normalized (lines~\ref{c16} to \ref{c17}).

\subsection{The distribution}

When computing the \emph{cascading survivability} for the nodes of a network, given a network and a specific $\alpha$, it is worth noting the utility of analysing the distribution of the $CS$ values, as previously illustrated for \emph{epidemic survivability} in Section~\ref{dist}.

By sorting the $CS$ values in descending order it is possible to compare different networks, according to a specific cascading failure scenario denoted by $\alpha$. Fig.~\ref{fig:CS} shows the $CS$ distribution of 15 of the networks considered in this work, in the case of a cascading failure with $\alpha=0.05$. It is interesting to note that most of the networks show a bimodal $CS$ distribution. This means that the nodes of such networks can be clearly divided in two groups: \emph{harmful} and \emph{not significant} in the case of a cascading failure. This behavior has been observed in other works such as \cite{journals/corr/abs-1006-4627}. Moreover, as observed, depending on the network the percentage of harmful nodes might vary. For instance, the \emph{fb4039} and the \emph{er400} networks start the distribution around $0.9$, however it is in the former where only a 5\% of the nodes represents a \emph{threat} in the case of cascading failures, while in the latter it is about 55\%. Finally, different types of complex networks show different $CS$ distribution curves, just like they show different $ES$ curves as represented in Section~\ref{dist}.

\section{Summary and Conclusions\label{sec:conclusions}}
In this paper we have proposed two new measures to evaluate the vulnerability of complex networks in two different dynamic multiple failure scenarios: epidemic-like and cascading failures.

Firstly, we have proposed a new network measure called \emph{epidemic survivability} ($ES$), which describes the vulnerability of each node of a network under a specific epidemic-like failure propagation scenario. Besides, a procedure to compute our novel measure has been provided. Sorting the $ES$ distribution of values of all nodes of a network in descending order, it is possible to analyze which nodes would be more vulnerable in the case of an epidemic failure. Furthermore, using this $ES$ distribution, network vulnerability can be compared for a specific epidemic scenario.

Secondly, we have presented a new network measure called \emph{cascading survivability} ($CS$), which characterizes how potentially dangerous a node is according to a specific cascading failure scenario. In addition, we have provided a procedure to compute $CS$. Then, as for the \emph{epidemic survivability} metric, we have noted the inherent usability related to the $CS$ distribution.

Lastly, we have computed $ES$ and $CS$ for the set of networks considered in this work, being each measure dependent on a specific failure scenario. Results have shown that distinct types of complex networks might react differently under the same dynamic multiple failure. In addition, results have revealed that a complex network might be more or less vulnerable, depending on the specific type of multiple failure scenario (i.e. epidemic-like or cascading failures). For instance, while the \emph{cogentco197} network shows a smooth decreasing curve of $ES$, the same network shows a bimodal distribution of $CS$, where about 25\% of nodes are not dangerous in the case of cascading failures.

To conclude, the \emph{methodology} that we have followed to evaluate the vulnerability of the nodes of a network in the case of dynamic multiple failures might be used in further investigations, considering other types of failures or models. This methodology is defined below:
\begin{enumerate}
	\item Define the set of networks to be analysed.
	\item Determine the failure scenario. 
	\item Choose a suitable model to simulate the failures.
	\item Define the value of all the parameters of the model.
	\item For each network, compute the vulnerability of the elements (e.g. nodes) of the network analytically or by performing simulations.  
\end{enumerate}

\section*{Acknowledgements}
This work is partially supported by Spanish Ministry of Science and Innovation projects TEC 2012-32336 and MTM 2011-27739-C04-03, and by the Generalitat de Catalunya research support program SGR-1202. This work is also partially supported by the Secretariat for Universities and Research (SUR) and the Ministry of Economy and Knowledge through AGAUR FI-DGR 2012 and BE-DGR 2012 grants (M.~M.)

\bibliography{references.bib}
\bibliographystyle{unsrt}

\end{document}